\documentclass[traditabstract]{aa} 
\usepackage{graphicx}
\usepackage{lscape}
\usepackage{natbib}
\usepackage{multirow}
\usepackage{ulem}
\usepackage{cancel}
\usepackage{soul}

\newcommand{\hi}{\mbox{H\,{\sc i}}} 
 
\newcommand{\hei}{\mbox{He\,{\sc i}}} 
\newcommand{\heii}{\mbox{He\,{\sc ii}}}

\newcommand{\feii}{\mbox{Fe\,{\sc ii}}}
\newcommand{\feiii}{\mbox{Fe\,{\sc iii}}}

\newcommand{\siii}{\mbox{Si\,{\sc ii}}}

\newcommand{\nv}{\mbox{N\,{\sc v}}}
\newcommand{\nii}{\mbox{N\,{\sc ii}}}
\newcommand{\niii}{\mbox{Ni\,{\sc ii}}}
\newcommand{\mgii}{\mbox{Mg\,{\sc ii}}}
\newcommand{\mgi}{\mbox{Mg\,{\sc i}}}
\newcommand{\cii}{\mbox{C\,{\sc ii}}} 
\newcommand{\crii}{\mbox{Cr\,{\sc ii}}} 
\newcommand{\znii}{\mbox{Zn\,{\sc ii}}} 
\newcommand{\civ}{\mbox{C\,{\sc iv}}} 

\def\grb{GRB\,080310}
\def\lsim{\mathrel{\hbox{\rlap{\lower.55ex \hbox {$\sim$}}\kern-.0em
\raise.4ex \hbox{$<$}}}} 
\def\gsim{\mathrel{\hbox{\rlap{\lower.55ex \hbox {$\sim$}}\kern-.0em
\raise.4ex \hbox{$>$}}}} 
\def\lya{Ly$\alpha$}
\def\lyb{Ly$\beta$}
\def\lyg{Ly$\gamma$}
\def\lyd{Ly$\delta$}
\def\lye{Ly$\epsilon$}
\def\ion#1#2{#1$\;${\small\rm\@Roman{#2}}\relax}

\def\kms{km s$^{-1}$}
\def\1star{$^{\star}$}
\def\2star{$^{\star\star}$}
\def\3star{$^{\star\star\star}$}
\def\4star{$^{\star\star\star\star}$}

\newcommand{\gpm}[3]{$#1^{+#2}_{-#3}$}
\newcommand{\swift}{{\it Swift}}

\newcommand{\Zsun}{\mbox{$Z_\odot$}}

\newcommand{\cmcube}{\mbox{cm$^{-3}$}}

\newcommand{\ergcmsHz}{\mbox{erg cm$^{-2}$ s$^{-1}$ Hz$^{-1}$}}

\newcommand{\percmsq}{cm$^{-2}$}
\newcommand{\persec}{s$^{-1}$}

\begin{document}

\title{Time-dependent excitation and ionization modelling of
  absorption-line variability due to \grb}


\author{P.~M. Vreeswijk\inst{1,2}
  \and
  C. Ledoux\inst{3}
  \and
  A.~J.~J. Raassen\inst{4,5}
  \and
  A. Smette\inst{3}
  \and
  A. De Cia\inst{1}
  \and
  P.~R. Wo\'zniak\inst{6}
  \and
  A.~J. Fox\inst{3,7,8}
  \and
  W.~T. Vestrand\inst{6}
  \and
  P. Jakobsson\inst{1}
}

\institute{
  Centre for Astrophysics and Cosmology, Science Institute, University
  of Iceland, Dunhagi 5, 107 Reykjavik, Iceland
  \email{pmvreeswijk@gmail.com}
  \and
  Dark Cosmology Centre, Niels Bohr Institute, University of
  Copenhagen, 2100 Copenhagen, Denmark
  \and
  European Southern Observatory, Alonso de C\'ordova 3107, Casilla
  19001, Santiago 19, Chile
  \and
  Astronomical Institute Anton Pannekoek, University of Amsterdam,
  Science Park 904, 1098 XH, Amsterdam, the Netherlands
  \and
  SRON Netherlands Institute for Space Research, Sorbonnelaan 2, 3584
  CA Utrecht, the Netherlands
  \and
  Los Alamos National Laboratory, MS-D466, Los Alamos, NM 87545, U.S.A.
  \and
  Institute of Astronomy, University of Cambridge, Madingley Road,
  Cambridge, CB3 0HA, United Kingdom
  \and
  Space Telescope Science Institute, 3700 San Martin Drive, Baltimore,
  MD 21218, U.S.A.
}


\abstract {We model the time-variable absorption of \feii{}, \feiii{},
  \siii{}, \cii{} and \crii{} detected in UVES spectra of \grb, with
  the afterglow radiation exciting and ionizing the interstellar
  medium in the host galaxy at a redshift of $z=2.42743$. To estimate
  the rest-frame afterglow brightness as a function of time, we use a
  combination of the optical $VRI$ photometry obtained by the RAPTOR-T
  telescope array -- which are presented in this paper -- and {\it
    Swift's} X-Ray Telescope observations.  Excitation alone, which
  has been successfully applied for a handful of other GRBs, fails to
  describe the observed column-density evolution in the case of
  \grb. Inclusion of ionization is required to explain the
  column-density decrease of all observed \feii{} levels (including
  the ground state $^6D_{9/2}$) and increase of the \feiii{} $^7S_3$
  level. The large population of ions in this latter level (up to 10\%
  of all \feiii{}) can only be explained through ionization of
  \feii{}, whereby a large fraction of the ionized \feii{} ions -- we
  calculate 31\% using the Flexible Atomic (FAC) and Cowan codes --
  initially populate the $^7S_3$ level of \feiii{} rather than the
  ground state. This channel for producing a significant \feiii{}
  $^7S_3$ level population may be relevant for other objects in which
  absorption lines from this level -- the UV34 triplet -- are
  observed, such as BAL quasars and $\eta$ Carinae. This provides
  conclusive evidence for time-variable ionization in the circumburst
  medium, which to date has not been convincingly detected. However,
  the best-fit distance of the neutral absorbing cloud to the GRB is
  200--400~pc, i.e. similar to GRB-absorber distance estimates for
  GRBs without any evidence for ionization. We find that the presence
  of time-varying ionization in \grb\ is likely due to a combination
  of the super-solar iron abundance ([Fe/H]=+0.2) and the low \hi{}
  column density (log $N(\hi{})=18.7$) in the host of \grb. Finally,
  the modelling provides indications for the presence of an additional
  cloud at 10--50~pc from the GRB with log $N(\hi{})\sim19$--20 before
  the burst and which became fully ionized by the radiation released
  during the first few tens of minutes after the GRB.}


\authorrunning{Vreeswijk et al.}
\titlerunning{Excitation and ionization modelling of
  time-variable absorption}

\maketitle

\section{Introduction}
\label{sec:intro}

Gamma-ray burst (GRB) afterglows can be detected at nearly any
wavelength up to very high redshifts
\citep{2009Natur.461.1254T,2011ApJ...736....7C} and are associated
with the deaths of massive stars \citep[for a recent review,
  see][]{2011arXiv1104.2274H}; they are therefore considered promising
probes of star formation at high redshift
\citep[e.g.][]{2000ApJ...536....1L}. However, in order to interpret
the wealth of information on the interstellar medium (ISM) of GRB host
galaxies gathered from GRB afterglow spectroscopy
\citep[e.g.][]{2007ApJ...666..267P,Fynbo09}, it is important to
understand in what way, and up to which distance, a GRB explosion is
affecting its host.

Several possible effects due to the brief but extremely powerful
radiation of a GRB and its afterglow have been predicted, such as the
gradual ionization of \hi{} and \mgii{}
\citep{1998ApJ...501..467P,2002ApJ...580..261P}, the excitation and
dissociation of H$_2$ molecules
\citep{2000ApJ...532..273D,2002ApJ...569..780D}, the destruction of
dust grains \citep{2000ApJ...537..796W,Fruchter01} and the
accompanying decrease in extinction and release of metals into the gas
phase \citep{2002ApJ...580..261P,2003ApJ...585..775P}. Apart from the
detection of excited H$_2$ molecules \citep{2009ApJ...701L..63S}, none
of these effects have been convincingly detected. For dust
destruction, this may be explained by the time scale being too short
(tens of seconds) for present observations to allow a firm detection.

One effect that was not predicted, but which has now been
unambiguously observed in several GRBs, is absorption-line variability
of fine-structure lines\footnote{The interaction of the total electron
  spin and the total electron angular momentum causes a fine-structure
  splitting of the atom levels, and the transitions with the lower
  energy levels corresponding to these excited levels are called
  fine-structure lines \citep[see][]{1968ApJ...152..701B}.} of ions
such as \feii\ and \niii\ \citep{2007A&A...468...83V,D'Elia09a}. This
variability has been shown to be due to the afterglow ultraviolet (UV)
photons exciting the neutral absorbers in the ISM at distances of a
hundred parsec up to well over a kiloparsec from the GRB
\citep{2006astro.ph..1057P,2007A&A...468...83V,D'Elia09a}.  These
distances are consistent with lower limit estimates for the neutral
gas ($>$ 50--100~pc) based on the presence of \mgi\ in the afterglow
spectra \citep{2006astro.ph..1057P}. They are also in agreement with
hydrodynamic calculations of the size of the {\it pre}-GRB ionization
bubble that is being created by the GRB progenitor star and its likely
cluster companions \citep{2008ApJ...682.1114W}.  Such a scenario, in
which the immediate environment is already mostly ionized by the time
that the GRB occurs, can also explain the difference between the
equivalent hydrogen column density measured from the soft X-rays,
$N$(H), and the neutral hydrogen column density, $N(\hi{})$, inferred
from \lya{} absorption in the optical/UV spectra
\citep{Watson07,Campana10,Schady11}.

In a companion paper \citep[][hereafter referred to as
  Paper~I]{DeCia}, we report in detail on our time-resolved
high-resolution spectroscopic observations of the \grb\ afterglow with
the Ultraviolet and Visual Echelle Spectrograph (UVES), mounted on the
Kueyen unit of ESO's Very Large Telescope (VLT). This sightline
displays an unusually low \hi{} column density at the GRB redshift
($z=2.42743$), log $N(\hi{})=18.7$, with an extreme iron and chromium
overabundance: ${\rm [Fe/H]}=+0.2$ and ${\rm [Cr/H]}=+0.7$ -- these
estimates include a correction for ionization effects. The values for
the carbon, oxygen and silicon abundances are instead rather typical
for GRB sightlines.  Another outstanding feature of the \grb\ UVES
spectra reported in Paper~I is the unique detection of the \feiii{}
UV34 triplet at 1895~\AA, 1914~\AA, and 1926~\AA, never seen before in
a GRB sightline. This, combined with the simultaneous decrease of the
column density population of all levels of \feii{}, including the
ground state, is suggestive of on-going ionization at the time the
UVES spectra were being secured.

In this follow-up paper we study this hypothesis in detail, by
modelling the column densities of \hi{}, \feii{}, \feiii{}, \siii{},
\cii{} and \crii{} observed in Paper~I as a function of time,
incorporating -- for the first time -- both photo-excitation and
-ionization in a consistent manner. An important input parameter for
our calculations is the afterglow brightness as a function of time,
which we estimate by combining the observed optical and X-ray
fluxes. The latter are derived from observations by the {\it Swift}
X-Ray Telescope, which are publicly available
\citep[see][]{Evans09}. For the optical we use the clear-filter and
$VRI$ light curves as measured by the RAPTOR-T array, which started
imaging the field as early as 32 seconds after the GRB trigger time
\citep[see][]{raptor8310gcn}; these data are also presented in this
paper.

This paper is organized as follows. We first describe the RAPTOR-T
measurements and present the broad-band $VRI$ and clear-filter light
curves in Sect.~\ref{sec:raptor}. In Sect.~\ref{sec:modelling} we
describe the implementation of the excitation and ionization processes
in our modelling code, where the reader is referred to
Appendix~\ref{sec:appendix} for the details.  We present the results
of our model fits to the \grb{} ionic column densities published in
Paper~I in Sect.~\ref{sec:results}. These results are discussed in
Sect.~\ref{sec:discussion} and we briefly summarize our findings in
Sect.~\ref{sec:conclusions}.

\section{The RAPTOR-T light curves}
\label{sec:raptor}

\begin{figure}[b]
  \includegraphics[width=9cm]{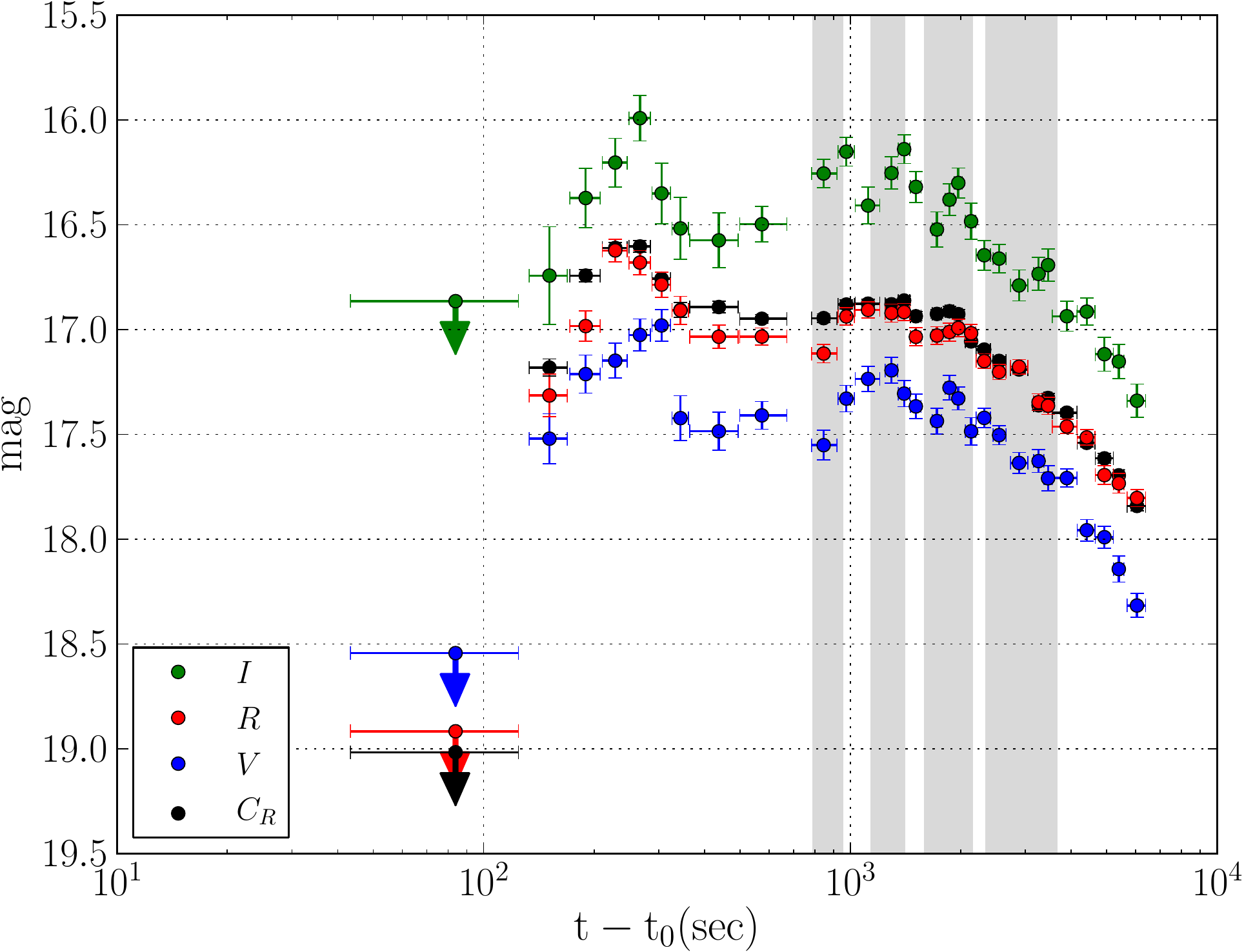}
  \caption{Light curves in $V$ (bottom), $R$, $I$ (top) and the clear
    filter as recorded by the RAPTOR-T telescope. The magnitudes have
    not been corrected for the Galactic foreground extinction. The
    four epochs at which the UVES spectra were taken are indicated
    with the grey vertical bands.}
  \label{fig:raptor}
\end{figure}

\begin{table*}
  \caption{Log of RAPTOR-T observations.}
  \label{tab:log}
  \centering          
  \begin{tabular}{r r r r c c c c}
    \hline\hline       
    T$_{\rm mid}$\tablefootmark{a} &
    T$_{\rm start}$\tablefootmark{a} &
    T$_{\rm end}$\tablefootmark{a} &
    Exp. Time &
    Clear filter\tablefootmark{b,c} &
    $V$\tablefootmark{b,c} &
    $R$\tablefootmark{b,c} &
    $I$\tablefootmark{b,c} \\
    (s) &
    (s) &
    (s) &
    (s) &
    (mag) &
    (mag) &
    (mag) &
    (mag) \\
    \hline                    
   83.8 &    43.3 &   124.4 &    40.0 &  $>$ 19.02 &  $>$ 18.54 &  $>$ 18.92 & $>$ 16.86 \\
  151.1 &   133.2 &   169.1 &    30.0 &   17.18 $\pm$   0.04 &    17.52 $\pm$   0.12 &    17.31 $\pm$   0.10 &    16.74 $\pm$   0.23  \\
  189.8 &   171.8 &   207.8 &    30.0 &   16.74 $\pm$   0.03 &    17.21 $\pm$   0.09 &    16.98 $\pm$   0.07 &    16.37 $\pm$   0.14  \\
  228.5 &   210.8 &   246.2 &    30.0 &   16.61 $\pm$   0.03 &    17.15 $\pm$   0.08 &    16.62 $\pm$   0.05 &    16.20 $\pm$   0.12  \\
  266.9 &   248.9 &   284.8 &    30.0 &   16.60 $\pm$   0.03 &    17.02 $\pm$   0.08 &    16.68 $\pm$   0.06 &    15.99 $\pm$   0.11  \\
  305.6 &   287.7 &   323.4 &    30.0 &   16.76 $\pm$   0.03 &    16.98 $\pm$   0.07 &    16.79 $\pm$   0.06 &    16.35 $\pm$   0.14  \\
  344.1 &   326.3 &   361.9 &    30.0 &   16.91 $\pm$   0.04 &    17.42 $\pm$   0.11 &    16.91 $\pm$   0.07 &    16.52 $\pm$   0.15  \\
  437.9 &   364.6 &   494.1 &    50.0 &   16.89 $\pm$   0.03 &    17.48 $\pm$   0.09 &    17.03 $\pm$   0.06 &    16.57 $\pm$   0.13  \\
  573.7 &   499.6 &   671.5 &    90.0 &   16.95 $\pm$   0.02 &    17.41 $\pm$   0.07 &    17.03 $\pm$   0.04 &    16.50 $\pm$   0.09  \\
  845.6 &   783.4 &   919.8 &    90.0 &   16.95 $\pm$   0.02 &    17.55 $\pm$   0.07 &    17.11 $\pm$   0.04 &    16.25 $\pm$   0.07  \\
  975.4 &   924.9 &  1025.8 &    90.0 &   16.88 $\pm$   0.02 &    17.33 $\pm$   0.06 &    16.94 $\pm$   0.04 &    16.15 $\pm$   0.07  \\
 1117.4 &  1031.7 &  1203.2 &    90.0 &   16.88 $\pm$   0.02 &    17.23 $\pm$   0.06 &    16.91 $\pm$   0.04 &    16.41 $\pm$   0.09  \\
 1294.8 &  1244.1 &  1345.4 &    90.0 &   16.88 $\pm$   0.02 &    17.19 $\pm$   0.06 &    16.92 $\pm$   0.04 &    16.25 $\pm$   0.08  \\
 1401.6 &  1350.9 &  1452.5 &    90.0 &   16.86 $\pm$   0.02 &    17.30 $\pm$   0.06 &    16.92 $\pm$   0.04 &    16.14 $\pm$   0.07  \\
 1508.7 &  1458.3 &  1559.2 &    90.0 &   16.94 $\pm$   0.02 &    17.37 $\pm$   0.06 &    17.03 $\pm$   0.04 &    16.32 $\pm$   0.07  \\
 1722.1 &  1671.7 &  1772.5 &    90.0 &   16.93 $\pm$   0.02 &    17.44 $\pm$   0.06 &    17.03 $\pm$   0.04 &    16.52 $\pm$   0.08  \\
 1863.5 &  1813.2 &  1913.6 &    90.0 &   16.91 $\pm$   0.02 &    17.28 $\pm$   0.06 &    17.01 $\pm$   0.04 &    16.38 $\pm$   0.07  \\
 1969.5 &  1919.2 &  2019.7 &    90.0 &   16.93 $\pm$   0.02 &    17.33 $\pm$   0.06 &    16.99 $\pm$   0.04 &    16.30 $\pm$   0.07  \\
 2134.5 &  2060.3 &  2196.7 &    90.0 &   17.06 $\pm$   0.02 &    17.48 $\pm$   0.07 &    17.02 $\pm$   0.04 &    16.48 $\pm$   0.09  \\
 2317.3 &  2202.2 &  2410.4 &   150.0 &   17.09 $\pm$   0.02 &    17.42 $\pm$   0.05 &    17.15 $\pm$   0.04 &    16.64 $\pm$   0.07  \\
 2544.2 &  2451.0 &  2658.5 &   150.0 &   17.15 $\pm$   0.02 &    17.50 $\pm$   0.05 &    17.20 $\pm$   0.04 &    16.66 $\pm$   0.07  \\
 2884.2 &  2734.5 &  3048.2 &   150.0 &   17.19 $\pm$   0.02 &    17.64 $\pm$   0.05 &    17.18 $\pm$   0.03 &    16.79 $\pm$   0.07  \\
 3260.5 &  3160.3 &  3368.0 &   150.0 &   17.36 $\pm$   0.02 &    17.63 $\pm$   0.05 &    17.35 $\pm$   0.04 &    16.73 $\pm$   0.08  \\
 3459.5 &  3373.6 &  3545.5 &   150.0 &   17.33 $\pm$   0.02 &    17.71 $\pm$   0.06 &    17.36 $\pm$   0.04 &    16.69 $\pm$   0.08  \\
 3892.5 &  3550.6 &  4149.2 &   300.0 &   17.40 $\pm$   0.02 &    17.71 $\pm$   0.04 &    17.46 $\pm$   0.03 &    16.94 $\pm$   0.07  \\
 4411.2 &  4154.6 &  4647.1 &   300.0 &   17.54 $\pm$   0.02 &    17.96 $\pm$   0.05 &    17.51 $\pm$   0.04 &    16.91 $\pm$   0.07  \\
 4929.8 &  4652.4 &  5214.3 &   300.0 &   17.61 $\pm$   0.02 &    17.99 $\pm$   0.05 &    17.69 $\pm$   0.04 &    17.12 $\pm$   0.08  \\
 5395.6 &  5220.7 &  5570.8 &   300.0 &   17.70 $\pm$   0.03 &    18.14 $\pm$   0.06 &    17.73 $\pm$   0.05 &    17.15 $\pm$   0.08  \\
 6047.1 &  5682.3 &  6385.4 &   420.0 &   17.84 $\pm$   0.02 &    18.32 $\pm$   0.06 &    17.80 $\pm$   0.04 &    17.34 $\pm$   0.08  \\
    \hline                  
  \end{tabular}
  \tablefoot{\scriptsize
    \tablefoottext{a}{Effective time at measurement midexposure,
      start and end, since the \swift\ BAT trigger on 10 March 2008 at
      08:37:58.65 UT.}
    \tablefoottext{b}{The magnitudes have not been
      corrected for the Galactic foreground extinction.}
    \tablefoottext{c}{The limiting magnitudes are $3\sigma$.}
  }
\end{table*}

GRB~080310 triggered the Burst Alert Telescope (BAT) onboard the
\swift\ satellite \citep{2008GCN..7382....1C} at 08:37:58.65 UT on
March 10, 2008. The RAPTOR-T telescope array began observing the BAT
position within 10.7 seconds after receiving the GCN alert, i.e. 32.4
seconds after trigger time. RAPTOR-T consists of four co-aligned 0.4-m
telescopes on a single fast-slewing mount and provides simultaneous
images in four photometric bands ($V$, $R$, $I$, and clear). The
system -- owned and operated by the Los Alamos National Laboratory
(LANL) -- is located at the Fenton Hill Observatory at an altitude of
2500~m in the Jemez Mountains of northern New Mexico. The RAPTOR-T
response sequence consists of 9, 20, and 170 exposures lasting,
correspondingly, 5, 10, and 30 seconds each, and separated by 5-second
intervals for readout. Approximately 25\% of individual frames were
rejected due to intermittent glitches in telescope tracking.

Aperture photometry was performed on co-added images using the
Sextractor package \citep{1996A&AS..117..393B} with object coordinates
fixed at the values measured on the reference image where the GRB is
detected at a high signal-to-noise ratio (S/N) in all four
channels. Instrumental light curves were then transformed to standard
Johnson magnitudes using SDSS photometry of stars in the vicinity of
the burst \citep{2008GCN..7396....1C} and equations of Lupton
(2005)\footnote{http://www.sdss.org/dr5/algorithms/sdssUBVRI
  Transform.html\#Lupton2005}. The results are listed in
Table~\ref{tab:log} and plotted in Fig.~\ref{fig:raptor}.

No optical emission was detected during the first two minutes after
the burst, down to a limiting magnitude of $R \sim 18.9$ ($3\sigma$).
The GRB is clearly detected in all subsequent co-adds starting at 133
seconds after the trigger.  Following a rapid increase in brightness
to a peak value at $R \simeq 16.6$ mag, the optical emission
fluctuates by a few tenths of a magnitude, and after $\sim30$ minutes
begins a slow decline. While the BAT light curve still shows a
detectable gamma-ray emission between 150 and 320 seconds after the
trigger \citep[see][]{2012arXiv1201.1292L}, the optical emission over
this time interval appears uncorrelated with the gamma rays. The
optical light curves from RAPTOR-T show no significant color
evolution. We determined the spectral slope $\beta$ (with $F_{\nu}
\propto \nu^{\beta}$) as a function of time, by fitting the $VRI$
magnitudes -- after correcting them for the Galactic foreground
extinction of $E_{B-V}=0.045$ \citep{1998ApJ...500..525S} -- at each
epoch. The resulting slope values do not show any trend in time, and
cluster around the value $\beta=-1.0$, with a standard deviation of
0.4 and an error in the mean of 0.07.

The RAPTOR-T measurements are generally consistent with those reported
in \citet{2012arXiv1201.1292L}. The RAPTOR $V$- and $R$-band magnitude
limits ($3\sigma$), at a mid-exposure time of 84~seconds after the
burst, correspond to $F_{V}<136~\mu$Jy and $F_{R}<78~\mu$Jy,
respectively. This $V$-band limit is in agreement with the {\it Swift}
$V$-band measurement of $F_{V}=247\pm140~\mu$Jy at the same epoch, but
our $R$-band limit is well below (almost a factor of three) the
prompt-emission fit featured in figure~10 of
\citet{2012arXiv1201.1292L}. Instead, it is fully consistent with the
alternative afterglow fit shown in their figure~8.

\section{Modelling the absorption-line variability}
\label{sec:modelling}

In Paper~I, we presented Voigt-profile fits to the aborption lines
detected in the \grb\ spectra, using four different velocity
components. Since these are very close in velocity ($<$ 60~\kms), it
is difficult to ascertain that they are indeed correctly separated in
the Voigt profile fit, even though a strong case can be made that
components ``b'' and ``c+d'' probably are. Moreover, the decomposition
is not unique, as additional components may be present that are hidden
in the profile. An added complication is that it is unclear which
fraction of the \hi{} column density belongs to which velocity
component, which is important for the modelling when ionizing
radiation is included. Inspection of the \feii{} and \feiii{} column
density evolution of the separate components indicates a generally
similar behaviour, which suggests that the components are at a
comparable distance. Preliminary modelling of the separate velocity
components ``b'' and ``c+d'' indeed results in distances that are the
same within the error margins. For these reasons, we have focused on
modelling the {\it total} column densities (listed in the last column
of table~3 in Paper~I) rather than those of the individual components.

\subsection{Photo-excitation}
\label{sec:photo-excitation}

Since the absorption-line variability observed in a handful of GRBs
can be generally well described by excitation of the host-galaxy ISM
by afterglow UV photons, we first set out to model the \grb\ observed
column density evolution as reported in Paper~I with photo-excitation
alone. The excitation fitting procedure applied here is similar to
that described in \citet{2007A&A...468...83V}, and which was also
applied by \citet{Ledoux09} and independently by
\citet{D'Elia09a,D'Elia09b,2010A&A...523A..36D}, but with two major
differences. First, we include the correct excitation flux \citep[see
  the erratum published by][]{2011A&A...532C...3V}, resulting in a
distance decrease of $\sqrt{4\pi}$ with respect to the old excitation
calculation. And second, instead of calculating the excitation at line
centre only, we now effectively integrate over the full line profile
to obtain the actual flux that is entering a particular layer. This
second change leads to a modest increase in the distance estimate of
about 10\%. The details of the excitation implementation are described
in Appendix~\ref{sec:appendix}.

For the \feii{} ion, we use the transition probabilities of the
371-level model atom as collected by \citet{Verner99}, including the
63 lower even-parity and 227 higher odd-parity levels. Note that the
transitions between even-parity levels are forbidden and have low
transition probabilities, while the transitions between even- and
odd-parity levels are electric dipole, i.e. allowed transitions.  The
observed resonance and fine-structure lines observed in the spectra of
\grb\ and other GRBs correspond to this latter group. This 371-level
model \feii{} atom is supplemented with transitions taken from the
Kurucz data base
\citep{1995all..book.....K}\footnote{http://kurucz.harvard.edu/atoms.html}. For
\feiii{}, we apply two different model atoms. One is calculated by
Raassen \&
Uylings\footnote{http://www.science.uva.nl/research/atom/levels/orth/iron},
from which we include 59 even-parity, and 214 odd-parity levels. The
alternative model, which we find to provide a slightly better fit,
combines the A-values for the forbidden transitions between the lowest
34 levels calculated by \citet{2010ApJ...718L.189B} with the allowed
transitions from \citet{2009ADNDT..95..184D}. In Paper~I we also
present measurements and upper limits of the excited-level column
densities of \siii{} and \cii{}, which we include in our fit as
well. The transition probabilities of both these ions are taken from
\citet{2003ApJS..149..205M} if present therein, and otherwise from the
NIST Atomic Spectra
Database\footnote{http://www.nist.gov/pml/data/asd.cfm} \citep[for the
  \siii{} and \cii{} transition probabilities and their references,
  see][]{kelleher:1285,wiese:1287}. We note that
\citet{2009A&A...508.1527B} have also calculated the A-values of
several \siii{} transitions; we find that using those values instead
of the NIST ones leads to a very similar amount of \siii{}
excitation. For \siii{} we include the ground level, $^2P^{\rm
  o}_{1/2}$, and its corresponding fine-structure level $^2P^{\rm
  o}_{3/2}$, and 19 higher even-parity levels. For \cii{}, we include
the lowest two levels ($^2P^{\rm o}_{1/2}$ and $^2P^{\rm o}_{3/2}$),
and 28 higher levels. For both these ions, the energy level of the
third odd-parity level is larger than the lower even-parity levels, so
the amount of atoms populating these other \siii{} and \cii{}
odd-parity levels is expected to be negligible. Finally, for \crii{},
we include the lowest 74 even-parity levels and nearly 400 odd-parity
levels, adopting the A-values of the forbidden transitions from
\citet{quinet1995}, and those of the allowed transitions from
\citet{1995all..book.....K}. For all ions, we have made sure that the
oscillator strengths (or equivalently, the transition probabilities)
of the relevant electric dipole transitions that are used to obtain
the ion column densities from the data through Voigt-profile fitting
(see Paper~I), are the same as used in the excitation modelling.

\subsubsection{Excitation modelling input flux spectrum}
\label{sec:inputflux_exc}

The afterglow UV flux at the GRB facing side of the absorbing cloud is
obtained by converting the $R$-band brightness as observed by the
RAPTOR-T telescope (see the previous Sect.~\ref{sec:raptor}) to the
host-galaxy redshift at a particular distance (a fit parameter) from
the GRB.  This conversion includes a correction for both the Galactic
extinction of $A_{\rm R}=0.12$~mag \citep{1998ApJ...500..525S}, and
any possible extinction in the host galaxy. The latter was found to be
of type Small Magellanic Cloud (SMC), with an estimated $V$-band
extinction (in the host-galaxy rest frame) of $A_{\rm
  V}=0.19\pm0.05$~mag \citep{2010ApJ...720.1513K}; we adopt this value
for our main fits, and assume that the dust responsible for this
extinction is located within the absorbing cloud. The $R$-band light
curve is interpolated in log space to obtain the brightness at any
given time to be used in the model calculations. To determine the flux
at different frequencies, we initially adopt a value for the spectral
slope of $\beta=-0.75$ \citep[the same as that adopted
  by][]{2012arXiv1201.1292L}. We note that $\beta$ represents the
intrinsic spectral slope, i.e. before it is affected by the
host-galaxy extinction (if non-zero).  The combination of this slope
with an extinction of $A_{\rm V}=0.19$~mag agrees well with the
observed spectral slope: $\beta=-1$, obtained from fitting the RAPTOR
$VRI$ data.  Apart from this default slope-extinction setting, we also
performed fits with zero host-galaxy extinction and the slope set to
the observed value of $\beta=-1$. Since the afterglow flux in the
X-ray regime is not relevant for excitation, we do not consider the
X-ray flux.

\subsubsection{Excitation-only fit result}
\label{sec:excitationfit}

\begin{figure}[t!]
  \includegraphics[width=9cm]{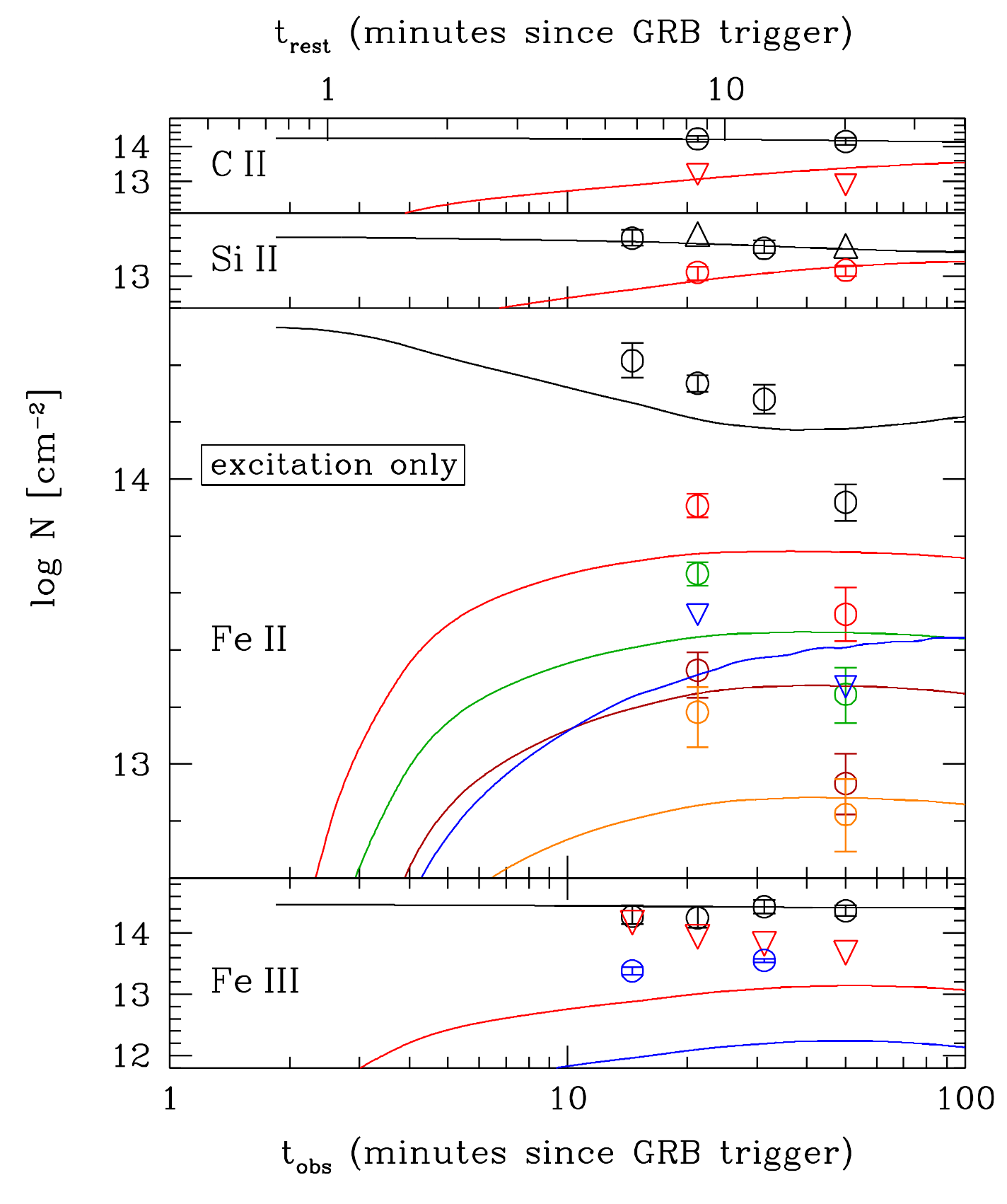}
  \caption{Photo-excitation modelling of the observed (total) column
    densities as a function of time, as measured in Paper~I (see their
    table~3), for \cii{}, \siii{}, \feii{}, and \feiii{}. The open
    circles are detections, while the open triangles indicate upper or
    lower limits ($3\sigma$).  The different colours denote the
    different ion levels: black for the ground state, and
    red-green-maroon-orange for the first four excited levels, while
    the \feii{} $^4F_{9/2}$ and \feiii{} $^7S_3$ levels are indicated
    in blue. The model fit describes the observed column densities
    very poorly, with a reduced chi-square of $\chi^2_{\nu}=21.3$.}
  \label{fig:excitation}
\end{figure}

The observed column density evolution of \feii{}, \feiii{}, \siii{}
and \cii{} (both ground-state and excited levels) are fit with an
excitation-only model, which includes the following fit parameters: 1)
the GRB to cloud distance, i.e. the distance from the GRB to the front
of the cloud, facing the GRB\footnote{Whenever we use the terms 'cloud
  distance', we refer to the distance from the GRB to the GRB-facing
  side of the absorbing cloud.}, 2) the linear cloud size, 3) the
pre-burst \feii{}, \feiii{}, \siii{} and \cii{} column densities, and
4) the Doppler parameter describing the velocity distribution of the
atoms.  The resulting fit, shown in Fig.~\ref{fig:excitation},
describes the observed column densities quite poorly, with a large
reduced chi-square value ($\chi^2_{\nu}=21.3$). One of the main
reasons for the poor fit is that all observed levels of \feii{} are
decreasing with time, which cannot be accommodated with excitation
alone. We note that -- based on the atomic transition probabilities
between the different levels -- it is not possible for a large
fraction of the pre-burst \feii{} atoms to be excited to levels above
those of the ground term ($^6D$). Moreover, transitions from these
higher levels are not observed (see Paper~I); e.g. see the \feii{}
$^4F_{9/2}$ level upper limits (indicated by the blue triangles) in
Fig.~\ref{fig:excitation}.  Another feature that is very difficult to
explain with excitation alone, is the very large observed fraction of
\feiii{} atoms in the excited $^7S_3$ level, of the order of 10\% (see
the blue level in the bottom panel of Fig.~\ref{fig:excitation}). This
level is severely underestimated by the model fit, despite the
best-fit cloud distance being lower than 50~pc.  For these reasons, we
can reject, with high confidence, the hypothesis that excitation alone
is responsible for the observed column density evolution along the
\grb{} sightline.

\subsection{Inclusion of photo-ionization}
\label{sec:photo-ionization}

\begin{figure*}[t]
  \includegraphics[width=9cm]{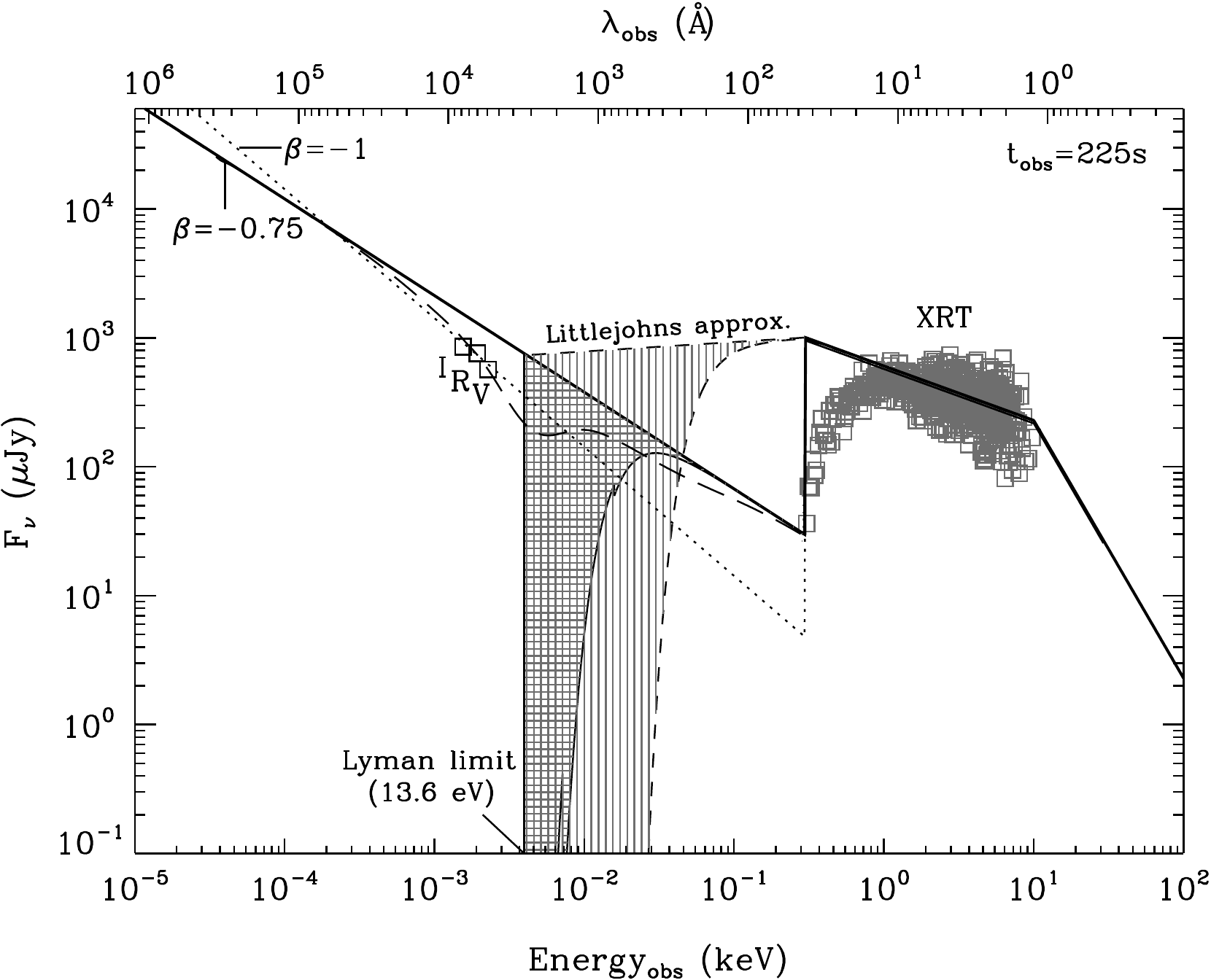}
  \includegraphics[width=9cm]{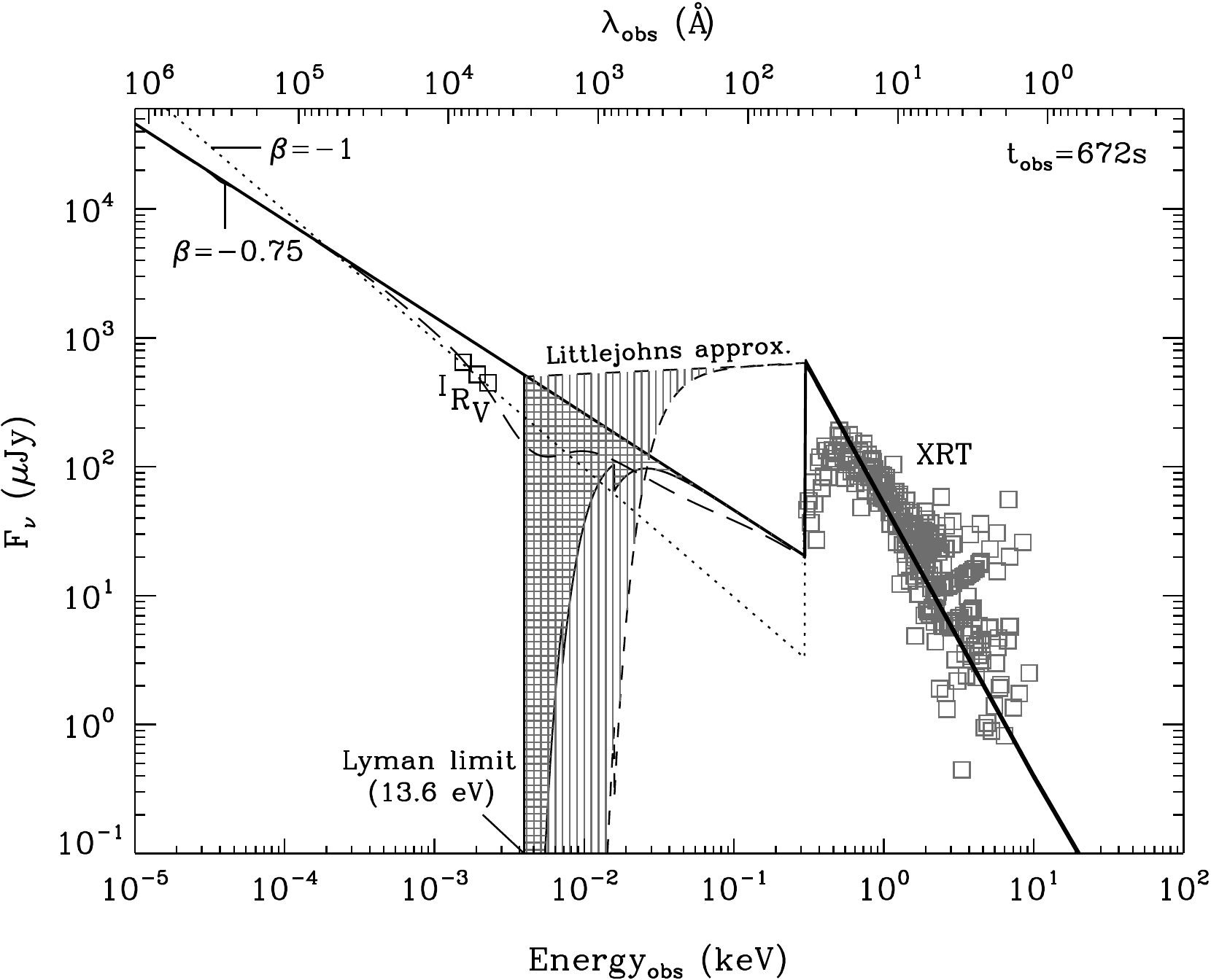}
  \caption{The adopted input flux spectrum that is arriving at the
    GRB-facing side of the observed cloud -- in observed flux units --
    depicted at two different epochs: t$_{\rm obs}=225$~seconds on the
    left, and t$_{\rm obs}=672$~seconds on the right. The solid line
    shows the default input spectrum with a spectral slope of
    $\beta=-0.75$ up to 0.3~keV, above which the X-ray flux (at
    1.73~keV) and spectral slope is adopted.  The RAPTOR $VRI$ and
    {\it Swift} XRT observations corresponding to these epochs are
    overplotted with open squares. The long-dashed line shows the
    default input spectrum modified by an extinction of $A_{\rm
      V}=0.19$~mag; since this extinction is placed inside the
    observed cloud, it is not observable at the front of the
    cloud. The dotted line shows the input spectrum assuming a
    spectral slope of $\beta=-1$, combined with no extinction. The
    dashed line between the observed optical and X-ray regimes shows
    our approximation to the Littlejohns model flux. The energy and
    flux limits of this figure are the same as those in figure~9 of
    \citet{2012arXiv1201.1292L} for easy comparison. The hashed
    regions show the flux decrease due to a foreground cloud with
    \hi{} column density log $N(\hi{})=18.9$ for the default input
    flux model (horizontal lines) and log $N(\hi{})=20.3$ for the
    alternative Littlejohns input spectrum (vertical lines) -- see
    Sect.~\ref{sec:results} and Table~\ref{tab:fitresults} for more
    details. The ionization edges of \hei{} at 24~eV/(1+$z$) and
    \heii{} at 54~eV/(1+$z$) can be spotted. We note that the spectral
    region in between the Lyman limit and the X-ray data is not
    constrained by imaging observations.}
  \label{fig:inputflux}
\end{figure*}

Excitation of an ion in a particular ionization state does not change
the total number of ions in that state. However, the observed \feii{}
column densities all clearly decrease in time, including the ground
state whose variability is generally not detected. This suggests that
\feii{} may be increasingly ionized (by the GRB afterglow) to higher
ionization states such as \feiii{} (see Paper~I).  This hypothesis is
supported by the detection of transitions of \feiii{}, involving both
the ground state and the $^7S_3$ excited level; this latter level has
never been observed before along a GRB sightline. Moreover, in Paper~I
we find that the \feiii{} $^7S_3$ level population is clearly
increasing with time. Given this observational evidence for
photo-ionization, and our finding above that photo-excitation alone
cannot reproduce the column density evolution observed, we have
included photo-ionization in our model calculations.

\subsubsection{Modelling input flux spectrum: inclusion of X-rays}
\label{sec:inputflux_ion}

With the inclusion of photo-ionization, we need to consider both the
UV and X-ray afterglow radiation.  X-ray photons photo-ionize species
such as \feii{}, \siii{} and \feiii{} mainly via the ejection of
inner-shell electrons. Since the X-ray flux for \grb\ is not a simple
extrapolation of the optical/UV flux with the optical spectral slope
\citep[see][]{2012arXiv1201.1292L}, we include the X-ray light curve
as measured by the \textit{Swift} X-ray Telescope (XRT). We retrieved
the 0.3--10 keV XRT afterglow light curve in count rate from the
\textit{Swift} repository \citep{Evans09} and separate it in 8
different time intervals (0--141, 141--185, 185--269, 269--393,
393--545, 545--615, 615--796, 796--7261~seconds after the trigger) in
order to limit the possible X-ray spectral evolution in each single
isolated light curve track. We then extract the spectrum from the
repository for each time interval and we convert the count rate light
curve to flux density accordingly. The monochromatic flux at 1.73 keV
(logarithmic average of the X-ray band) was calculated assuming the
correspondent spectral slopes for each window.  This X-ray light curve
replaces the $R$-band extrapolation in the regime above 0.3~keV (in
the observer's frame, corresponding to 1.0~keV in the host galaxy rest
frame). In the region 0.3--10~keV we adopt the spectral slopes
determined for the different time intervals, and beyond 10~keV we
adopt a spectral slope of $\beta=-2$ at all times
\citep[see][]{2012arXiv1201.1292L}. The X-ray spectra and assumed
spectral slopes are shown for two time intervals (185--269 and
615--796~seconds) in Fig.~\ref{fig:inputflux}.

We also performed fits with an alternative to the input flux spectrum
described above. This alternative is motivated by modelling of the
\grb\ afterglow by \citet{2012arXiv1201.1292L}, which suggests that
the early-time flux (up to about 1800~s in the observer's frame), in
between roughly 3~eV and 300~eV, is much higher than the $\beta=-0.75$
(or $\beta=-1$) extrapolation from the optical \citep[see figure~9
  of][]{2012arXiv1201.1292L}. We note that this regime has no
observations that are able to constrain the proposed model.  The
Littlejohns model flux is approximated by interpolation of the RAPTOR
optical and {\it Swift} X-ray light curves in between
3000~\AA\ (3.6~eV) and 300~eV (both in the observer's frame). Below
and above this region the flux used is the same as the original input
flux spectrum described above. In Fig.~\ref{fig:inputflux} we show the
default input spectrum (solid line) and the Littlejohns alternative
(dashed line), at two different epochs. In the modelling, the input
spectrum is constructed by interpolation of the the RAPTOR $R$-band
and {\it Swift} XRT light curves for each new time step.

\subsubsection{Cross section of \feii{} ionization to different
  levels of \feiii{}}
\label{sec:sigmafe23}

Our program incorporates well-known astrophysical processes
\citep[e.g.][]{2006agna.book.....O}, and we refer the reader to
Appendix~\ref{sec:appendix} for a detailed description of how the
photons, with the input flux described above, that are travelling
through the cloud, are exciting and ionizing the ions therein.  We
stress that ionization is taken into account for all relevant ions,
i.e. \hi{}, \hei{}, \heii{}, \feii{}, \feiii{}, \siii{}, \cii{} and
\crii{}, and that we properly take into account the fraction of
\feii{} that will be ionized to \feiii{} (rather than to higher
ionization states), as calculated by \citet{Kaastra93} for the
different ion shells. Excitation is included for all ions except for
hydrogen and helium.  One very important non-standard aspect -- the
calculation of the cross section of \feii{} ionization to different
(excited) levels of \feiii{} -- is discussed here.

When \feii{} is ionized to \feiii{}, the \feiii{} ion will not
necessarily be in its ground state, at least not immediately.  We have
calculated the photo-ionization cross section from \feii{} to specific
levels of \feiii{}, using two different codes: the suite of programs
developed by \citet{1981tass.book.....C} and the Flexible Atomic Code
(FAC) developed by \citet{2003ApJ...582.1241G,gu2004}. The Cowan code
is a self consistent Hartree-Fock model with relativistic
corrections. The FAC package is also a self-consistent program, which
models the wave functions to self-consistency by including the electron
screening. Relativistic effects are taken into account by means of the
Dirac Coulomb Hamiltonian.

The lowest configurations in \feii{}, $3d^7$ and $3d^6 4s$, strongly
overlap, having $3d^6(^5D) 4s$ $^6D_{9/2}$ as its ground
state. However the $^6D$ magnetic J-sublevels are just slightly higher
in energy and are all populated (see Paper~I). Since no absorption
features from higher lying levels in \feii{} have been observed (see
table~3 of Paper~I), we have focused on ionization from the low lying
$^6D$~ levels. The character of the configuration that the ground
state belongs to ($3d^6 4s$) results in a photo-ionization process
that is both complex and interesting. There are several channels to
ionization from the ground configuration: {\bf (a)} $3d^6 4s$
$\rightarrow$ $3d^6$ by ionizing the outer 4s-electron to the
p-continuum by means of a photon absorption, and {\bf (b)} $3d^6 4s$
$\rightarrow$ $3d^5 4s$ by ionizing the 3d-electron to the p- or
f-continuum by means of a photon absorption.

The result of {\bf (a)} is the population of the \feiii{} $3d^6$ $^5D$
states and {\bf (b)} will end up in the \feiii{} $3d^5(^6S) 4s$ $^7S$
or $^5S$ states. In \feiii{} there is quite an energy difference
(3.7~eV) between $3d^6$ $^5D$ and $3d^5(^6S) 4s ^7S$ . The fact that
absorption features arising from these states are observed and not
from states in between, indicates that photo-ionization plays the
dominant role rather than collisional excitation.

\begin{figure}[t!]
  \includegraphics[width=8.6cm]{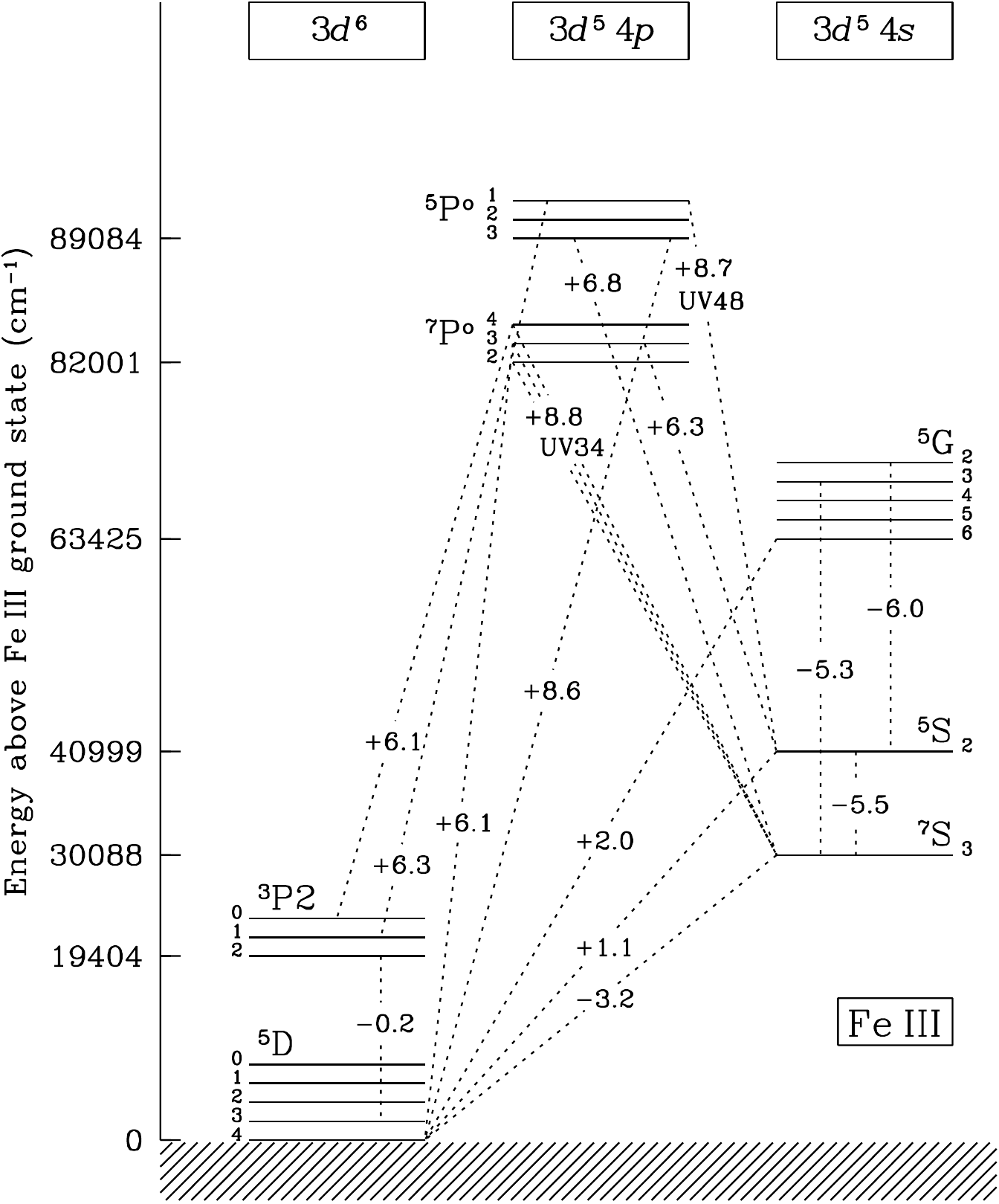}
  \caption{Partial energy level (or Grotrian) diagram for the relevant
    lower terms of the three lowest configurations of \feiii{}:
    $3d^6$, $3d^5 4s$ and $3d^5 4p$ (indicated at the top). The
    horizontal solid lines depict the energy levels, labelled with the
    term and $J$-value. Selected transitions are shown with dotted
    lines between levels. For each transition we list the logarithm of
    the transition probability (or Einstein A-coefficient, in
    [\persec]) of the strongest transition between the terms, adopting
    the Raassen \& Uylings values (see
    Sect.~\ref{sec:photo-excitation}).}\label{fig:energydiagram}
\end{figure}

In the approach using the Cowan Code the even configurations $3d^7$
and $3d^6 4s$ were applied and the odd continuum states $3d^6$
$\epsilon p$ and $3d^5 4s$ $\epsilon p$ and $\epsilon f$.  In case of
FAC the \feii{} $3d^6 4s$ and $3d^7$, and \feiii{} $3d^5 4s$ and
$3d^6$ configurations were introduced in the modelling. Comparison of
the results from the Cowan and FAC programs shows a very good general
agreement. Table~\ref{tab:sigmafe23} lists the calculated cross
sections for the relevant levels of \feiii{} and the corresponding
fraction of \feii{} ionizations that populate that particular \feiii{}
level. These numbers are used directly in our modelling
program.\footnote{We note that before we had calculated these numbers,
  we included the fraction of \feii{} ionizations that populate the
  $^7S_3$ level of \feiii{} as a free parameter in our model fit, with
  a resulting best-fit value of 30--35\%.} We find that only a small
fraction (9\%) will directly populate the \feiii{} ground term, while
31\% of the new \feiii{} ions will in fact populate the $^7S_3$ level.
The majority -- 57\% -- will populate the levels of the $3d^5(^4G) 4s$
$^5G$ term. However, these latter will quickly decay to the $3d^6$
$^5D$ ground term.  Fig.~\ref{fig:energydiagram} shows a partial
energy diagram of some relevant terms of \feiii{}.  We note that many
more terms exist, but are not depicted for clarity. For each relevant
transition (indicated with the dotted line), we list the logarithm of
the transition probability. E.g. for the strongest transition between
the $3d^5 4s$ $^5G$ and $3d^6$ $^5D$ terms, this is
$A=10^{+2.0}$~\persec.  The reciprocal of this number provides the
time in seconds in which the ions in the upper level would decay to
the lower level in the absence of radiation. 

\begin{table}[t]
  \caption{Cross sections for ionization 
    of ground-term \feii{} ions to different excited levels of \feiii{}, 
    as calculated with the FAC and Cowan codes.}
  \label{tab:sigmafe23}      
  \begin{tabular}{rrr}
    \hline\hline
    \feiii{} level\tablefootmark{a} & 
    cross section &
    fraction of total \\
    &
     ($\times10^{-19}$~\percmsq) &
    \% \\
    \hline
    $3d^6$ $^5D_4$ (1) &  3.50 &  1.96\\
    $^5D_3$ (2) &  3.48 &  1.95\\
    $^5D_2$ (3) &  3.47 &  1.95\\
    $^5D_1$ (4) &  3.46 &  1.94\\
    $^5D_0$ (5) &  1.88 &  1.05\\
    \hline
    $3d^5(6S)4s$ $^7S_3$ (18) & 55.74 & 31.26\\
    $^5S_2$ (26) &  1.69 &  0.95\\
    \hline
    $3d^5(4G)4s$ $^5$G$_6$ (35) & 18.54 & 10.40\\
    $^5G_5$ (36) & 19.03 & 10.67\\
    $^5G_4$ (37) & 20.01 & 11.22\\
    $^5G_3$ (38) & 22.49 & 12.61\\
    $^5G_2$ (39) & 25.01 & 14.03\\
    \hline
  \end{tabular}
  \tablefoot{\scriptsize \tablefoottext{a}{The \feiii{} level is
      indicated with the configuration, the term and subscript J value
      (and the level number -- ordered in energy -- starting from
      the ground level).}  }
\end{table}

\subsubsection{Comparison with ionization calculations in the literature}

As a consistency check, we compared the amount of ionization computed
in our program with calculations in the literature. Since our
ionization model does not take into account
recombination,\footnote{This is not required, because the relevant
  time scale of our calculations -- hours to days after the GRB --
  is negligible compared to the recombination time scale at typical
  ISM densities.} while most calculations in the literature do, the
comparison options are limited to GRB ionization studies. Examples of
these are the studies of \citet[][]{2002ApJ...580..261P},
\citet{2003ApJ...585..775P} and \citet{2002ApJ...569..780D}, in which
not only the ionization induced by the GRB is calculated, but also the
accompanying destruction of dust and dissociation of H$_2$. Since our
program does not include dust destruction, this makes a comparison
with these calculations difficult.
However, we were able to compare our program with the H, He, and N
photo-ionization calculations by \citet{Prochaska08b}, and find
consistent results. We use their Eq.~8 for the GRB\,050730 afterglow
luminosity over the same time span $t_{\rm obs}=10$--1000~s and adopt
their set-up with $n_{\rm H}=10$~\cmcube, a nitrogen-to-hydrogen
abundance ratio of $10^{-6}$ (roughly 0.01 solar metallicity) and
assume that before the GRB all the ions are in the singly ionized
state. We then switch on the GRB~050730 afterglow and
follow the progressive ionization of \nii{} to higher ionization
states, and find that after 1000~s, the \nv{} column density remaining
is log $N(\nv{})=13.8$, compared to their log $N(\nv{})=14$. Also the
ionization structure at $t_{\rm obs}=1000$~s computed by our program
is very similar to that depicted in their figure 3.

\subsubsection{$\chi^2$ Minimization and fit parameters}

The model column densities computed by our program, as detailed in
Appendix~\ref{sec:appendix}, are fit to the observed \grb{} column
densities at their respective epochs. We use the Fortran 90 version of
the MINPACK {\tt lmdif} $\chi^2$ minimization routine
\citep{minpack,minpackuserguide}, which is based on the
Levenberg-Marquardt method. We have made this program parallel with
OpenMP, so it can be run faster on a shared-memory computer
cluster. The formal errors of the fit parameters are estimated by
computing the co-variance matrix, and taking the square root of the
diagonal elements.

The fit parameters are the same as those used in the excitation-only
case described at the end of Sect.~\ref{sec:photo-excitation}: the GRB
to cloud distance (i.e. the distance from the GRB to the GRB-facing
side of the cloud), the cloud size, the Doppler parameter $b$, and a
pre-burst column density for each ion included in the fit.  We again
initially fix the spectral slope to $\beta=-0.75$
\citep[see][]{2012arXiv1201.1292L}, in combination with a host-galaxy
extinction of $A_{\rm V}=0.19$~mag \citep{2010ApJ...720.1513K}, but
also experiment with the combination $\beta=-1.0$ and zero
extinction. The ions included are: \feii{}, \feiii{}, \siii{}, \cii{}
and \crii{} (apart from \hi{}, \hei{} and \heii{}, see below), i.e.
all low-ionization species with a total column density measurement at
one or more epochs as reported in table~3 of Paper~I. For all of
these, except \crii{}, we include excitation.

\begin{figure}[t!]
  \includegraphics[width=9cm]{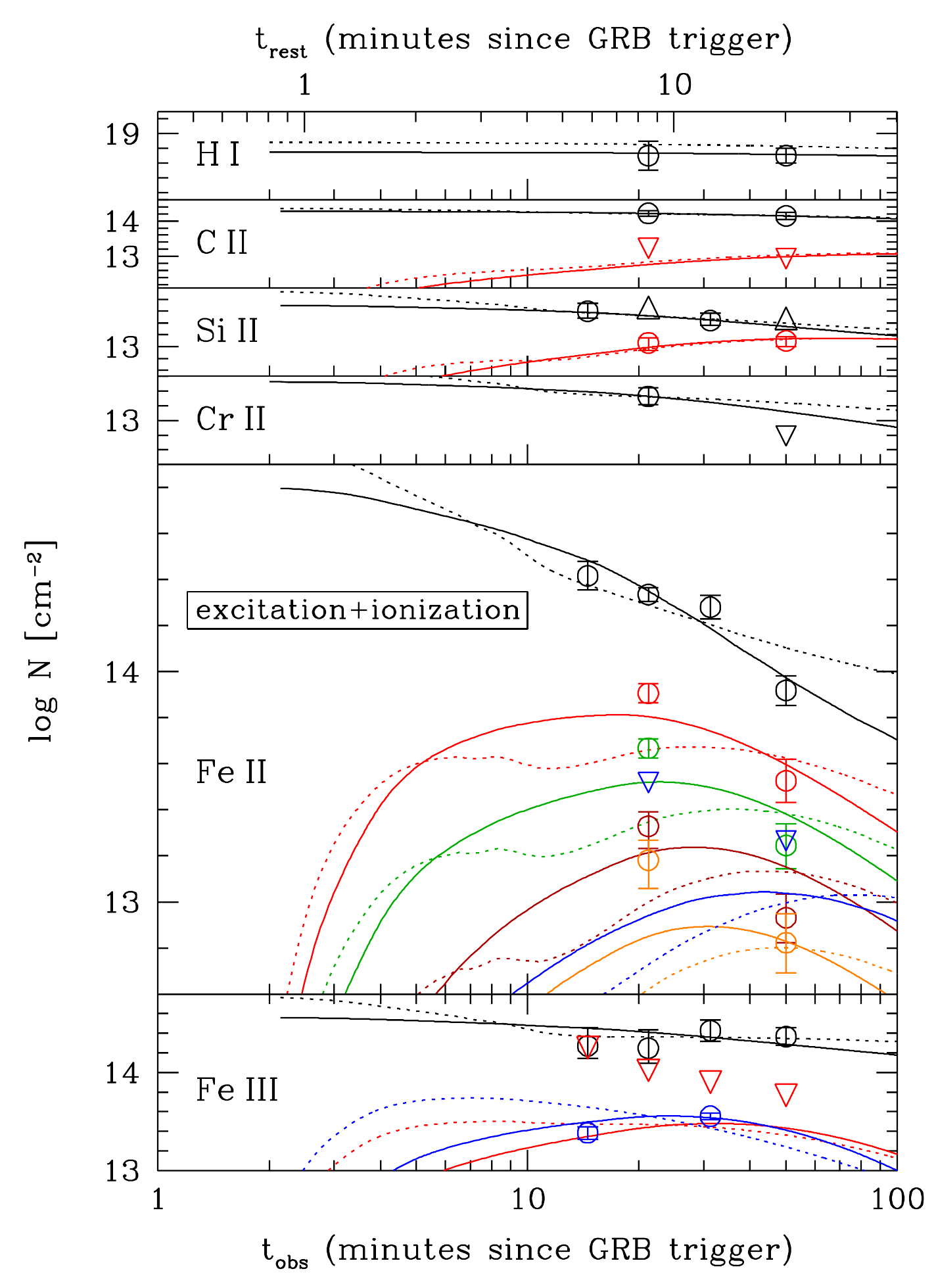}
  \caption{Photo-excitation and -ionization modelling of the observed
    (total) column densities as a function of time, as measured by
    \citet{DeCia} (see table~3 of Paper~I), for -- from top to bottom
    panels -- \hi{}, \cii{}, \siii{}, \crii{}, \feii{}, and \feiii{}.
    The solid and dotted lines correspond to the best-fitting model
    assuming the default and the Littlejohns input flux, respectively.
    The different colours of the symbols and lines have the same
    meaning as in Fig.~\ref{fig:excitation}. Although overall the
    model provides a reasonable description of the column density
    evolution, the observed excited levels of \feii{} at epoch II are
    significantly underestimated. See the text and
    Table~\ref{tab:fitresults} for more details.}
  \label{fig:modelfit}
\end{figure}
\begin{figure}[t!]
  \includegraphics[width=9cm]{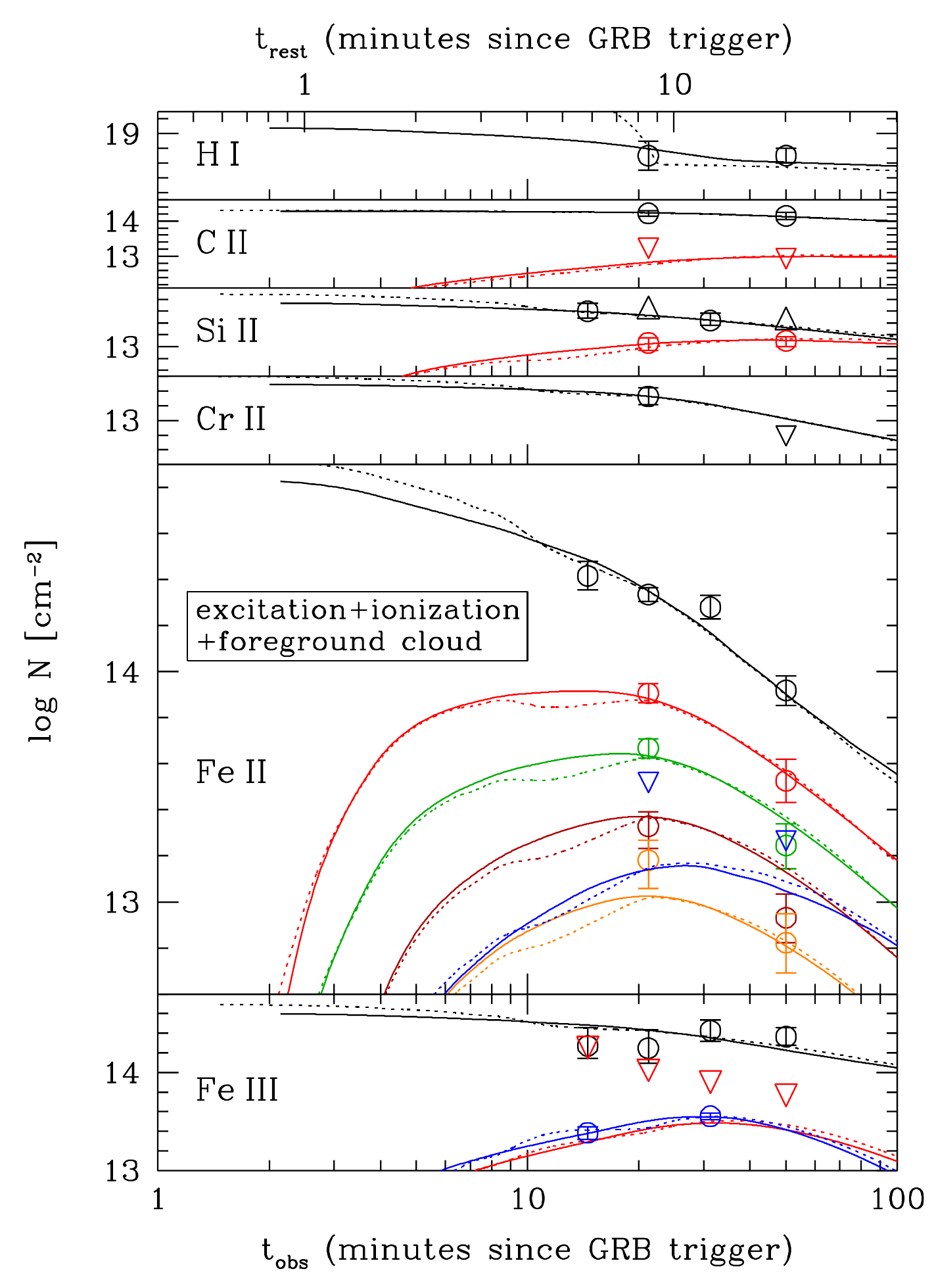}
  \caption{Same as Fig.~\ref{fig:modelfit}, but including a foreground
    cloud situated in between the GRB and the observed absorber.  The
    solid and dotted lines correspond to the best-fitting model
    assuming the default and Littlejohns input flux, respectively. The
    foreground absorber is mostly ionized by the time of the 2nd epoch
    UVES spectrum and can therefore escape a clear detection. See the
    text and Table~\ref{tab:fitresults} for more details.}
  \label{fig:addcl_modelfit}
\end{figure}

As discussed in Paper~I, the velocity profiles of \feii{} and \feiii{}
are markedly different from those of \siii{} and \cii{}.  The former
are dominated by component ``b'' at $-$20~\kms\ from the systematic
velocity and with a Doppler broadening parameter value of
$b=13$~\kms. However, a considerable column density is contained in
the other components ``a'', ``c'' and ``d'' as well, leading to an
overall broad velocity structure for \feii{} and \feiii{}. In
contrast, the vast majority (although not all) of the \siii{} and
\cii{} column densities is located in the narrow ``c'' component at
the systematic velocity, with $b=7$~\kms. For this reason, we split
the velocity broadening fit parameter into two: one for \feii{} and
\feiii{}: $b_{\scriptsize\rm \feii{},\feiii{}}$, and one for \siii{} and \cii{}:
$b_{\scriptsize\rm \siii{},\cii{}}$.

In Paper~I, we also constrained log $N(\hi{})$ to 18.7 at two
different epochs.  This \hi{} column density is an important quantity,
as it can -- if it is sufficiently large -- effectively shield the
low-ionization species (such as \feii{} and \siii{}) from the ionizing
photons.  Besides \hi{}, we also include \hei{} and \heii{}, which are
also important for shielding, albeit at higher photon energies
(starting from 24~eV). These helium ions do not require additional fit
parameters, as we fix the \hei{} column density at the solar abundance
value \citep[i.e. 8.5\% of N(\hi{}),][]{Asplund09} and set the
pre-burst \heii{} column density to zero. We note that the inclusion
of a significantly larger amount of \hei{} -- a possibility if there
is a large column density of pre-burst {\it ionized} hydrogen -- does
not affect our results. If the \hei{} abundance is included in the fit
as a free parameter, the best-fit value is consistent with the adopted
value: [\hei{}/\hi{}]$=(12\pm17)$\%.

\section{Results}
\label{sec:results}

The resulting fit to the total column densities is shown in
Fig.~\ref{fig:modelfit}. The solid (dotted) lines correspond to the
best-fitting model assuming the default (Littlejohns) input flux
discussed in Sect.~\ref{sec:photo-ionization}. The goodness-of-fit and
best-fit parameter values are listed in the first column of
Table~\ref{tab:fitresults}. The model fit in which the Littlejohns
input flux is adopted is very poor, with $\chi^2_{\nu}=8.34$, and we
therefore discard it without listing the unreliable best-fit parameter
values in Table~\ref{tab:fitresults}. The quality of the model fit
that assumes the default input flux is reasonable, with
$\chi^2_{\nu}=2.51$.  This rather high value for the reduced
chi-square seems to be mainly caused by the model underpredicting the
observed population of the \feii{} excited levels.  Assuming a
negligible host-galaxy extinction ($A_{\rm V}=0$~mag), combined with
the observed spectral slope of $\beta=-1$, leads to a slightly
improved fit with a chi-square value of $\chi^2_{\nu}=2.36$, but with
resulting best-fit values consistent within the errors of the default
fit (with $\beta=-0.75$ and $A_{\rm V}=0.19$~mag).

\begin{table}[t]
  \centering          
  \caption{Column-density evolution modelling fit results.\tablefootmark{a}}
  \label{tab:fitresults}      
  \begin{tabular}{llll}
    \hline\hline
    Incl. foreground cloud?        & no  & yes & yes \\
    Incl. Littlejohns flux?        & no  & no  & yes \\
    \hline
    $\chi^2_{\nu}$ (degrees of freedom) & 2.51 (17) & 1.50 (14) & 1.69 (14) \\
    Cloud distance\tablefootmark{b} (pc)            & $363\pm86$ & $235\pm97$  & $260\pm107$ \\  
    Cloud size (pc)                & $0\pm149$  & $126\pm141$ & $55\pm146$ \\
    $b_{\scriptsize\rm \feii{},\feiii{}}$ (\kms) & 50\tablefootmark{c} & 50\tablefootmark{c} & $38\pm13$ \\
    $b_{\scriptsize\rm \siii{},\cii{}}$ (\kms)   & $2.1\pm0.7$ & $1.5\pm0.6$ & $1.4\pm0.5$ \\  
    log $N(\hi{})$ [\percmsq]        & \gpm{18.75}{0.08}{0.10} & \gpm{18.64}{0.12}{0.17}& \gpm{18.60}{0.11}{0.14} \\
    log $N(\feii{})$ [\percmsq]      & \gpm{14.80}{0.03}{0.03} & \gpm{14.83}{0.04}{0.04}& \gpm{14.93}{0.07}{0.08} \\
    log $N(\feiii{})$ [\percmsq]     & \gpm{14.56}{0.08}{0.10} & \gpm{14.60}{0.08}{0.10}& \gpm{14.70}{0.10}{0.13} \\
    log $N(\siii{})$ [\percmsq]      & \gpm{13.70}{0.06}{0.07} & \gpm{13.74}{0.07}{0.08}& \gpm{13.88}{0.07}{0.09} \\
    log $N(\cii{})$ [\percmsq]       & \gpm{14.28}{0.06}{0.08} & \gpm{14.28}{0.06}{0.08}& \gpm{14.30}{0.07}{0.08} \\
    log $N(\crii{})$ [\percmsq]      & \gpm{13.52}{0.12}{0.16} & \gpm{13.49}{0.20}{0.37}& \gpm{13.60}{0.13}{0.18} \\
    FC\tablefootmark{d} distance\tablefootmark{b} (pc)             & & 50\tablefootmark{c} & $12\pm8$ \\
    FC\tablefootmark{d} size (pc)                 & & $39\pm715$  & $9\pm19$ \\
    FC\tablefootmark{d} log $N(\hi{})$ [\percmsq]   & & \gpm{18.9}{0.7}{1.0} & \gpm{20.30}{0.15}{0.24} \\
    \hline                    
  \end{tabular}
  \tablefoot{\scriptsize \tablefoottext{a}{The spectral slope and
      host-galaxy extinction were fixed to the values: $\beta=-0.75$
      and $A_{\rm V}=0.19$~mag, respectively.}
    \tablefoottext{b}{This is the distance from the GRB to the
      GRB-facing side of the cloud.}
    \tablefoottext{c}{Maximum allowed fit value reached.}
    \tablefoottext{d}{Foreground Cloud.}}
\end{table}

Table~\ref{tab:fitresults} shows that the best-fit \siii{} and \cii{}
Doppler broadening parameter is very low: $b_{\scriptsize\rm
  \siii{},\cii{}}=2.1\pm0.7$~\kms.  As we discussed in the previous
section, the observed $b$-parameter value is low as well:
$b_{\scriptsize\rm \siii{},\cii{}}=7$~\kms. To investigate this modest
discrepancy further, we also ran a model in which only \hi{}, \hei{},
\siii{} and \cii{} are included, i.e. without \feii{}, \feiii{} and
\crii{}, and with the $b$-parameter fixed to the observed value of
$b=7$~\kms. Although the resulting distance to the GRB-facing side of
the cloud is very small, less than 50~pc, the cloud size becomes more
than a kiloparsec, i.e. the average distance is quite large. Indeed,
forcing the cloud to be very compact, with a cloud size fixed at 1~pc,
the best-fit distance, both without and with the Littlejohns input
flux, is roughly 600~pc. These results indicate that the majority of
\siii{} and \cii{} ions might be at a different location (further away
from the GRB) or spread out over a larger region than the bulk of the
\feii{} and \feiii{} ions. This is supported by the very different
velocity profiles that these ions display (see Paper~I). However, for
other GRB sightlines for which a cloud distance has been determined
independently for \feii{} and \siii{} excitation
\citep[e.g.][]{2010A&A...523A..36D,2011arXiv1108.1084D}, the best-fit
distances are consistent. This suggests that -- as one would expect --
the \feii{} and \siii{} atoms are probably located at comparable
distances from the GRB.

Since the \feii{} excited levels are under-predicted by the model, we
attempted to place an additional cloud along the line of sight, in
between the GRB and the observed absorber or cloud. If the additional
cloud would be sufficiently close to the burst, it would become
completely ionized during the first few tens of minutes (in the
observer's frame) of the arrival of the GRB radiation, and would not
reveal itself in the observations. But at the same time it would
partially shield the observed cloud from ionizing radiation released
during the first minutes after the GRB, allowing the observed cloud to
be closer to the burst, increasing the amount of excitation. This
scenario could only work in case the two clouds have a velocity offset
(already 10--20~\kms\ is sufficient, which is not unlikely), to avoid
the observed cloud to be in the absorption-line shadow of the
foreground cloud.  The fit with such an additional cloud is shown in
Fig.~\ref{fig:addcl_modelfit}, again with the solid (dotted) curves
corresponding to the model fit adopting the default (Littlejohns)
input flux, and the best-fit parameter values are listed in the 2nd
and 3rd column of Table~\ref{tab:fitresults}. The addition of such a
foreground cloud results in a lower value for the chi-square:
$\chi^2_{\nu}=1.50$ (assuming the default input flux), but at the
expense of three additional fit parameters: the distance, size and
column density of the foreground cloud (FC, see
Table~\ref{tab:fitresults}). An F-test suggests that the fit
improvement introduced by the foreground cloud is significant,
providing $F=\frac{(\chi^2-\chi_{\rm FC}^2)/(\nu - \nu_{\rm
    FC})}{\chi_{\rm FC}^2/\nu_{\rm FC}}=4.8$ (where $\nu$ is the
number of degrees of freedom) and a null probability of $P<0.005$;
i.e. there is less than 0.5\% chance that such an improvement is
random.

We also ran model fits with both a foreground cloud and adopting the
alternative Littlejohns input flux. As described in
Sect.~\ref{sec:photo-ionization}, this input flux is much higher (up
to a factor of 10) than the default input flux in between 0.3~eV and
300~eV \citep[see][]{2012arXiv1201.1292L}, i.e. leading to much more
ionizing radiation. As already mentioned above, a model fit with the
Littlejohns input flux {\it without} an additional cloud describes the
observed column density evolution very poorly. However, an additional
cloud with a neutral hydrogen column density almost that of a damped
\lya\ (DLA) system at 10--20~pc from the GRB, is capable of absorbing
most of the extra ionizing radiation, leading to a very reasonable fit
(with $\chi^2_{\nu}=1.71$). The best-fit parameter values for this
model are listed in the 3rd column of Table~\ref{tab:fitresults}, and
the resulting column-density evolution is shown with a dotted line in
Fig.~\ref{fig:addcl_modelfit}.

\section{Discussion}
\label{sec:discussion}

The evolution of the \feii{} and \feiii{} column densities observed at
the \grb\ redshift (see Paper~I, and Sects.~\ref{sec:photo-ionization}
and \ref{sec:results}), combined with our modelling, clearly shows
that ionization of \feii{} is taking place. A very strong argument in
favour of ionization and a vital ingredient for the modelling, is that
according to our calculations (see Sect.\ref{sec:sigmafe23}) a large
fraction (31\%) of \feii{} ionizations will -- initially -- populate
the \feiii{} $^7S_3$ level.  Without taking this effect into account,
we found it impossible to explain the large fraction ($\sim$10\%) of
\feiii{} that is observed to be in this particular level (see
Paper~I). This channel for producing a significant \feiii{} $^7S_3$
level population may be relevant for other objects in which absorption
lines from this level -- the UV34 triplet -- are also observed, such
as broad absorption line (BAL) quasars and $\eta$ Carinae.  As it
takes about 1000~seconds for the \feiii{} $^7S_3$ level population to
decay spontaneously down to the ground term, the \feii{} ionization
rate needs to be significant at this time scale for this process to be
relevant. In BAL quasars, the UV48 triplet, at 2062, 2068, and
2079~\AA, is sometimes detected. The lower energy level from which the
UV48 triplet arises, $^5S_2$ (see Fig.~\ref{fig:energydiagram}),
receives only a small fraction of the \feii{} ions that are ionized to
\feiii{} (1\%, see Table~\ref{tab:sigmafe23}), and so these lines are
expected to be much weaker than the UV34 triplet. We checked for the
presence of these UV48 absorption lines in the UVES spectra of \grb,
and indeed do not detect them. In the sample of unusual BAL quasars of
\citet{Hall02}, the detection of the UV34 triplet is much more common
than UV48, which would be expected in case ionization of \feii{} is
the dominant mode of populating the UV34 lower level. However, in case
the UV48 absorption is stronger than that of UV34, such as in SDSS
2215--0045 \citep{Hall02,2012arXiv1204.3629V}, the above \feiii{}
excited-level population scenario, which works well for \grb, does not
provide a viable explanation.

Time-variation of \hi{} and metal-column densities due to the ongoing
ionization by the GRB and afterglow radiation has been predicted
\citep[e.g.][]{1998ApJ...501..467P}, but has never been convincingly
detected before \citep[see][]{Thone11}. This applies not only to
neutral-medium ions such as \hi{} and \feii{}, but also to
high-ionization species as \civ{} and \nv{}
\citep{Prochaska08b,Fox08}. The reason that ongoing photo-ionization
is observed for \grb\ is not that the observed neutral material along
the \grb\ sightline is much closer to the GRB than in other cases. We
find a distance range of 200--400~pc (depending on the adopted input
flux and the inclusion or not of a foreground cloud, see
Sect.~\ref{sec:results} and below), while other GRBs for which only
excitation was detected have distance estimates as low as 50~pc
\citep[][]{2011arXiv1108.1084D}. In Table~\ref{tab:distances} we have
collected the GRB-cloud distance estimates from the literature,
allowing for a direct comparison with the distance estimate for
\grb. We note that, in the absence of a foreground cloud, a lower
limit of about 100~pc can be placed on the GRB-absorber distance, by
just considering the non-variation in the \hi{} column density between
21 and 50 minutes post-burst.  If the absorber would have been much
closer, we would have detected a significant \hi{} column-density
change.

\begin{table*}[ht]
  \centering
  \caption{GRB absorber distances, as of April 2012.}
  \begin{tabular}{lccccccccl}
    \hline \hline
    GRB & Instrument & $z$ & Distance & Size & P$^a$ &log $N$(\hi{}) & [X/H] & $X$ & Ref. \\
    &            &     &  (pc)    & (pc) &       &               &       &     &    \\
    \hline
    020813 & LRIS+UVES & 1.25 & 50--100 &     &       &               &       &   &1, 2\\
    \textbf{050730} & UVES & 3.97 &$124\pm20^b$ & \gpm{147}{68}{54}& E &$22.10\pm0.10$ & $-2.18\pm0.11$ & S & 3, 4\\
    051111 & HIRES & 1.55 & a few hundred    &                                 & E &    &  &    &5, 6\\
    \textbf{060418} & UVES & 1.49 &$480\pm56$& & E   &    $>21.0$    & $<-0.5$ & Zn &7, 4\\
    \textbf{080310} & UVES & 2.43 & 200--400 & 0--200 & E+I    &$18.70\pm0.10$  & $-1.2\pm0.2$ &Si&8, 9 \\
    \textbf{080319B} &UVES& 0.94 & 560--1700 & & E &  & & &10, 4\\
    080330 &UVES & 1.51&\gpm{79}{11}{14} & & E & & & &11, 4\\
    081008 &UVES+FORS & 1.97&$52\pm6^c$ & & E &$21.11\pm0.10$ & $-0.87\pm0.10$ & Si&12\\
    090426 & LRIS+FORS& 2.61 & $\gsim80^d$ & & I &\gpm{18.7}{0.1}{0.2} &    & & 13\\
    090926 & X-shooter& 2.11    & $677\pm42^e$ & & E &$21.60\pm0.07$ & $-1.85\pm0.10$& S & 14, 4\\
    \hline
  \end{tabular}
  \tablefoot{\scriptsize The distances derived from a
    photo-excitation/photo-ionization model of the column density
    variability, based on high-resolution spectroscopy, highlighted in
    bold, are considered to be more reliable. The excitation distances
    have been corrected for $\sqrt{4\pi}$ according to
    \citet{2011A&A...532C...3V}.\\
    \tablefoottext{a}{Process modelled:
      photo-excitation (E) or photo-ionization (I).}
    \tablefoottext{b}{A former analysis of the Magellan Clay/MIKE
      echelle spectrum \citep{Chen05} suggested a cloud distance
      $d<100$ pc \citep{2006astro.ph..1057P}.}
    \tablefoottext{c}{Component I (Component II lies at
      $200^{+60}_{-80}$ pc).} 
     \tablefoottext{d}{We consider the 090426 \lya\ variation
       detection to be marginal, and conservatively list this distance
       estimate as a lower limit.}
    \tablefoottext{e}{Main component (the
      second lies at $\sim5$~kpc).}
  }
  \tablebib{\scriptsize (1) \citet{Dessauges-Zavadsky06}; (2)
    \citet{Savaglio04}; (3) \citet{Ledoux09}; (4)
    \citet{2011A&A...532C...3V}; (5) \citet{2006ApJ...646..358P}; (6)
    \citet{2006astro.ph..1057P}; (7) \citet{2007A&A...468...83V}; (8)
    This work; (9) Paper~I: \citet{DeCia}; (10) \citet{D'Elia09a};
    (11) \citet{D'Elia09b}; (12) \citet{2011arXiv1108.1084D}; (13)
    \citet{Thone11}; (14) \citet{2010A&A...523A..36D}}
  \label{tab:distances}
\end{table*} 

We investigated whether the very low \hi{} column density or
super-solar iron abundance along the \grb\ sightline is the reason for
the unique detection of ongoing ionization. We did so by running
models with most parameters fixed to the best-fitting model (with
$\chi^2_{\nu}=1.5$ in Table~\ref{tab:fitresults}), but varying the
\hi{} column density and iron abundance, to see how these affect the
number of ions detected in the \feiii{} $^7S_3$ level. As we have
shown above, a significant population of this excited level is a clear
sign of ongoing ionization of \feii{}. Table~\ref{tab:max7S3} shows
the expected peak \feiii{} $^7S_3$ column density as a function of
different \hi{} column densities (rows) and iron abundances
(columns). In the column with fixed $N_{\rm FeII,FeIII}$, we fixed the
pre-burst \feii{} and \feiii{} column densities at their best-fit
values (middle column) of Table~\ref{tab:fitresults} for the four
different \hi{} column densities, while in the last two columns, all
the iron was assumed to be in the singly ionized state before the GRB
exploded; we note that this latter assumption is only valid at higher
\hi{} column densities (log $N(\hi{})\gsim20$).

  \begin{table}[h!]
    \centering
    \caption{Maximum \feiii{} $^7S_3$ column density reached as a
      function of the assumed \hi{} column and iron abundance in the
      \grb\ absorber.}
    \label{tab:max7S3}      
    \begin{tabular}{llll}
      \hline\hline
      log $N(\hi{})$ & fixed $N_{\rm FeII,FeIII}^a$ & [Fe/H]$=-1.0^b$
      & [Fe/H]$=+0.2^{b,c}$ \\
      \hline
      18.6 & 13.55 & 11.83 & 13.02 \\
      19.6 & 13.24 & 12.52 & 13.71 \\
      20.6 & 12.69 & 12.95 & 14.14 \\
      21.6 & 12.05 & 13.20 & 14.37 \\      
      \hline                    
    \end{tabular}
    \tablefoot{\scriptsize \tablefoottext{a}{In this column the
        \feii{} and \feiii{} column densities were fixed at the
        best-fit values of Table~\ref{tab:fitresults}, i.e. for higher
        \hi{} column densities the iron abundance effectively decrease
        with respect to the log $N(\hi{})=18.6$ case.}
      \tablefoottext{b}{For these runs at fixed iron abundance, we
        have assumed that before the onset of the GRB, [Fe/H] =
        [\feii{}/\hi{}], i.e. all the iron ions are in the singly
        ionized state and all the hydrogen is neutral, which is only a
        good approximation at higher \hi{} column densities.}
      \tablefoottext{c}{The iron abundance for \grb\ was determined to
        be [Fe/H]=+0.2 when including ionization corrections (see
        Paper~I).}  }
\end{table}

Considering log $N(\feiii ^7S_3)=13$ to be the approximate lower limit
for a clear detection of this level in the UVES spectra,
Table~\ref{tab:max7S3} shows that increasing the \hi{} column density
by a factor of about 100 or more, while fixing the \feii{} and
\feiii{} column densities at the best-fit values of
Table~\ref{tab:fitresults}, would have resulted in a non-detection of
the \feiii{} excited level. This is due to the increased shielding of
the low-ionization metals from the ionizing radiation by the \hi{} and
\hei{} atoms. However, when fixing the abundance at the observed value
for iron along the \grb\ sightline: [Fe/H]=+0.2 (see Paper~I), the
\feiii{} excited level is detected at any \hi{} column density. At a
more typical iron abundance for GRB sightlines, [Fe/H]$=-1.0$, the
\feiii{} UV34 triplet is detectable only at the higher \hi{} column
density end. We note that a column density of log $N(\hi{})=21.6$ at
0.1~\Zsun\ implies a considerable \feii{} column: log
$N(\feii{})=16.1$, and this increases by at least a factor of 10 when
assuming [Fe/H]$=+0.2$. At such large \feii{} columns, dust is not
unlikely to be present. The presence of dust would complicate the UV34
triplet detection at high \hi{} columns. Dust obscuration would not
only decrease the amount of ionization taking place, but would also
make it more difficult to detect a bright afterglow - required to
secure high-quality spectra - in the first place. Therefore, the
reason for the unique detection of the \feiii{} UV34 triplet in the
\grb\ spectra appears to be a combination of the super-solar iron
abundance and the low \hi{} column along this sightline. This ensures
the presence of a sufficient amount of iron, while at the same time
avoiding too much \hi{} and \hei{} shielding and dust obscuration.

If the detection of \feii{} ionization is indeed due to a combination
of the super-solar iron abundance and the low \hi{} column density
along the \grb\ sightline in the host, then the (non-)detection of the
\feiii{} UV34 triplet can be used to put constraints on the \hi{}
column density along a GRB sightline with an \feii{} detection in case
it cannot be inferred from the spectrum. The latter is the case at
$z\lsim2$, when \lya\ is not redshifted enough to be included in the
optical wavelength range of spectrographs on ground-based
telescopes. The strength of the \feiii{} UV34 triplet, however,
depends on various quantities besides the iron abundance and \hi{}
column, such as the GRB-absorber distance, the afterglow peak
luminosity and brightness evolution, and the time at which the spectra
are taken. It is therefore difficult to provide a simple scaling
relation between the \hi{} column, iron abundance and UV34 triplet
strength.

But for GRBs for which most of the above quantities can be constrained
through absorption-line photo-excitation modelling, it is possible to
determine a lower limit on the \hi{} column density from the \feiii{}
UV34 triplet non-detection. As our team has already performed such
modelling on GRB~060418 \citep[at
  $z=1.490$,][]{2007A&A...468...83V,2011A&A...532C...3V}, we can
readily determine this limit on \hi{} for this sightline. The UV34
triplet is not detected in the GRB~060418 UVES spectra, with a
$3\sigma$ upper limit on the rest-frame equivalent width (column
density) of 0.03~\AA\ (log $N(\hi{})=12.6$). Modelling the excitation
and ionization with our code, in which we vary the \hi{} column
density, we find that this UV34 detection limit corresponds to an
\hi{} column density limit of log $N(\hi{})>21.0$. Using the total
zinc column density measured for this sightline \citep[log
  $N(\znii{})=13.09\pm0.01$,][]{2007A&A...468...83V}, and assuming
that most of the zinc is in the singly ionized state, the \hi{} column
density lower limit derived above implies an upper limit on the
metallicity of [Zn/H]$<-0.5$. Determining these \hi{} column-density
and corresponding metallicity limits for the entire sample of
Table~\ref{tab:distances} requires photo-excitation and -ionization
modelling for each sightline separately, which is out of the scope of
the current paper.

Our simplest model, in which the GRB afterglow is ionizing and
exciting a cloud at a distance of about 360~pc, does not provide a
satisfactory description of the observations. As can be seen in
Fig.~\ref{fig:modelfit}, the model underestimates the ground-term
fine-structure level population. One potential reason for this lack of
\feii{} excitation -- or abundance of ionization -- may be that
additional neutral material is present in between the GRB and the
absorber responsible for the absorption features observed in the
spectra. This additional absorber needs to be ionized by the time that
the first couple of spectra were taken, as otherwise it would reveal
itself in the observed spectra.  Placing such an additional cloud
closer to the GRB, with log $N(\hi{})\sim19$ at a distance of tens of
parsecs, improves the model fit significantly, as shown by the solid
curves in Fig.~\ref{fig:addcl_modelfit}. Also in the case where the
Littlejohns input flux is adopted (depicted by the dotted curves in
Fig.~\ref{fig:addcl_modelfit}), the model with a foreground cloud
provides a very reasonable description of the observed column-density
evolution of the different ionic species. In this case, the foreground
cloud is required to have a higher neutral hydrogen column density
(log $N(\hi{})=20.3$) and to be closer to the GRB (12~pc), in order to
be able to absorb the additional ionizing photons in the Littlejohns
input flux. The similar chi-squares for the additional-cloud model
using the default and Littlejohns input fluxes does not allow us to
favour one input flux over the other; however, in the model without an
additional cloud, the default input flux is clearly favoured.

We tested if a log $N(\hi{})=20.3$ cloud at 12~pc, with an assumed
metallicity of one tenth of solar and using the Littlejohns input
flux, would imply an observable \nv{} variation
\citep[see][]{Prochaska08b,Fox08} in our spectra.  In Paper~I we
report a constant \nv{} column density: log $N(\nv{})=14.10\pm0.04$
and log $N(\nv{})=14.05\pm0.02$ at epochs II and IV, respectively.  In
this test, we adopt a metallicity of one-tenth of the solar abundance,
and we assume that all the nitrogen is singly ionized before the
burst. We find that the \nii{} ions are very quickly ionized to higher
ionization states: at 6 minutes after the burst (observer's frame) the
\nv{} column density in the foreground cloud is already below log
$N(\nv{})=13$, and by the time of the 1st epoch spectrum (13 minutes
after the burst), practically all the nitrogen has been ionized to
states higher than \nv{}. Also, if the foreground cloud is indeed
ionized within about ten minutes of the arrival of the first GRB
photons, it is very difficult to infer its presence in sightlines
where only ongoing excitation is observed.

Although the introduction of an additional cloud is a rather ad-hoc
solution for improving the model fit of Fig.~\ref{fig:modelfit} --
mitigating the ionization with respect to the amount of excitation --
the existence of an additional cloud in the vicinity of the burst is
not unexpected, as GRBs are thought to occur in gas-rich massive-star
forming regions \citep[e.g.][]{2007ApJ...666..267P}.  We note that the
presence of such an additional cloud is consistent with the
host-galaxy $N({\rm H})$-equivalent X-ray absorption as inferred from
the {\it Swift} XRT data \citep[$\log N(\mbox{H})=21.7\pm0.05$ and
  $\log N(\mbox{H})<21$ -- assuming solar metallicity -- for the
  time-averaged averaged Windowed Timing and Photon Counting modes,
  respectively,][]{Evans09}. Thus, although the presence of a
foreground cloud is plausible, we cannot exclude a different origin
for the underestimate of the \feii{} excitation (or overestimate of
ionization) in our default model fit.

\section{Conclusions}
\label{sec:conclusions}

We modelled the variability of the ionic column densities of various
species (including \hi{}, \hei{}, \heii{}, \feii{}, \feiii{}, \siii{},
\cii{} and \crii{}) in the circumburst medium of \grb\ \citep[reported
  in a companion paper by:][]{DeCia} with a photo-excitation and
-ionization radiative transfer code. The rest-frame near-infrared to
X-ray spectrum of the afterglow radiation and its time evolution, an
important input parameter in the modelling, is inferred by combining
the RAPTOR-T $VRI$ light curves -- also presented in this paper -- and
the X-ray light curve as observed by {\it Swift}.  We find that
excitation alone, which has been successfully applied to other GRBs,
is not able to explain the \grb\ observations; ionization is clearly
required. The strongest evidence for ionization is presented by the
clear detection of the UV34 triplet of \feiii{} from the lower level
$^7S_3$. The large fraction of \feiii{} ions measured to be in this
level, 10\%, can only be explained through ionization of \feii{}; we
calculate that 31\% of all \feii{} ions that end up as \feiii{} will
first populate this $^7S_3$ level. This is the first conclusive
evidence for the detection of time-variable photo-ionization induced
by a GRB afterglow.

Despite this evidence for photo-ionization, the distance between the
GRB and the absorbing medium that we infer (200--400~pc) is very
similar to that in other GRB sightlines for which such a distance
estimate was possible. We find that the main reason for detecting
time-variable ionization in this GRB and not in others is the
super-solar iron abundance ([Fe/H]=+0.2) in combination with the low
\hi{} column density (log $N(\hi{})=18.7\pm0.1$) along this sightline.

The combined photo-excitation and -ionization modelling provides
tentative evidence for the presence of an additional absorbing cloud,
with log $N(\hi{})\sim19$--20, at a distance of 10--50~pc from the
GRB, even though this cloud is almost completely ionized by the
afterglow within a few tens of minutes (in the observer's frame) of
the arrival of the GRB radiation.  Future time-resolved
high-resolution spectroscopic observations of low-\hi{} GRB sightlines
could provide additional constraints on the existence of pre-burst
neutral gas in the GRB vicinity.

\begin{acknowledgements}
  PMV is grateful for the support from the ESO Scientific Visitor
  program in Santiago, Chile. PRW and WTV acknowledge support for the
  RAPTOR and Thinking Telescopes projects from the Laboratory Directed
  Research and Development (LDRD) program at LANL. ADC acknowledges
  support from the ESO DGDF 2009, 2010 and the University of Iceland
  Research Fund. PJ acknowledges support by a Project Grant from the
  Icelandic Research Fund. The Dark Cosmology Centre is funded by the
  Danish National Research Foundation. The modelling performed in this
  paper was mostly performed on the excellent computing facilities
  provided by the Danish Centre for Scientific Computing (DCSC). We
  kindly thank Gudlaugur Johannesson for use of his 24-core work
  station when the DCSC servers would be down, and Peter Laursen for
  the insightful discussions on \lya{} scattering.  Last but not
  least, we are grateful for the professional assistance of the VLT
  staff astronomers, in particular Claudio Melo and Dominique Naef,
  who secured the UVES observations on which this paper is based.
\end{acknowledgements}

\bibliographystyle{aa} 
\bibliography{references}

\appendix
\section{Details of the time-dependent photo-excitation and -ionization calculations}
\label{sec:appendix}


This appendix describes in detail our time-dependent photo-excitation
and -ionization calculations of the neutral medium nearby the GRB,
along the line-of-sight. Although this includes well-known
astrophysical processes, it allows for a transparent comparison with
similar future studies.

Our program is rather basic when compared to a photo-ionization code
such as CLOUDY. It does not include recombination, which is a
reasonable assumption due to the very short time scale (of the order
of hours to a day) that the GRB afterglow is bright and that
high-resolution spectra can be secured. With an approximate rate of
10$^{-13}$~cm$^3$~s$^{-1}$, the recombination time scale at typical
ISM densities is orders of magnitude larger. Second, our 1D
calculations are performed along the line-of-sight only and we do not
take into account afterglow photons that have scattered off particles
elsewhere in the absorbing medium and into the sightline. And third,
we consider the source of afterglow photons to be very small compared
to the distance from the source to the absorbing medium.  However, the
important asset of our code is its ability to take in a time-variable
input source, and calculate the resulting column density evolution,
for both the ground state as well as excited levels, as a function of
time.

Photo-excitation and -ionization are treated separately but
self-consistently in our model. Photo-excitation involves bound-bound
transitions in the ion, while photo-ionization involves bound-free
transitions; hence the exciting photons have quasi-discrete
wavelengths, while the ionizing photons constitute a continuum with
photon energies larger than the ionization threshold. Therefore, for
the excitation we use a flux array that is calculated at specific
wavelengths only -- the central wavelengths of the relevant
transitions -- while for ionization we use a continuum flux
array. This continuum array starts at the lowest ionization threshold
of the ions used in the calculation (typically \hi{} at 13.6~eV) up to
100~keV, with logarithmic increments to have most of the wavelength
resolution around the lower photon energies. Ideally, we would use a
single continuum flux array, and calculate the imprint of both
excitation and ionization as the photons move into the absorbing
cloud, but this would be very CPU intensive as such an array would
require both a large wavelength range, from the far-infrared to hard
X-rays, as well as a sufficient resolution over this range. The two
flux arrays overlap, but over a fairly short wavelength
range. E.g. excitation of \feii{} ions is due to photons at particular
wavelengths up to the ionization potential of \feii{}, i.e. 16.2~eV,
and therefore they are in the range of the ionizing flux array. In
this overlapping region, we take into account the flux decrease due to
ionization on the exciting flux array.

The absorbing cloud is divided into a number of plane-parallel
layers. The initial (pre-burst) column densities of the cloud are
spread out evenly between the layers such that each layer is composed
of sublayers of pure \hi{}, \hei{}, \heii{}, and any additional ion
such as \feii{}, \feiii{}, and \siii{}. The number of layers is
determined by our requirement that each layer is optically thin ($\tau
< 0.05$) both at the \hi{} ionization threshold, as well as at the
central wavelength of any excitation transition. To mimic a continuous
excitation and ionization proces, we also require not more than 5\% of
the ions in any sublayer to be excited or ionized in any iteration, or
time step. This -- together with the amount of flux arriving at the
cloud front (see Sect.~\ref{sec:photo-excitation} and below) --
determines the time step, $\Delta t$, of the calculation.

The flux, $F_{\nu}$ (in \ergcmsHz), is known at any given time and
distance from the GRB. For the spectral range from the infrared to
300~eV, we use the RAPTOR-T $R$-band light curve and adopt a spectral
slope of $\beta=-0.75$ or $\beta=-1$, depending on the assumed amount
of host-galaxy extinction (see below). Above 300~eV we use the {\it
  Swift} X-ray light curve, as described in
Sect.~\ref{sec:photo-ionization}. The observed fluxes are converted to
the host-galaxy rest frame. For each time step, starting at 25~s in
the rest frame of GRB~080310, up to the epoch of the last UVES
spectrum, the corresponding flux is propagated through the cloud. It
decreases as $F_{\rm nlayer}=F_{\rm nlayer-1} \, e^{-\tau}$ -- with
$\tau$ being the optical depth -- as in every (sub-)layer photons are
exciting and ionizing ions. The propagating flux is also affected by
the dust extinction in the absorption medium, which is fixed to
$A_{\rm V}=0.19$~mag \citep{2010ApJ...720.1513K} when adopting a
spectral slope of $\beta=-0.75$, or to zero in combination with
$\beta=-1$. The extinction at wavelengths other than at the centre of
the $V$-band filter is calculated using the analytic fits of
\citet{Pei92}, assuming an SMC-type extinction.

For the excitation optical depth, we use:

\begin{equation}
  \tau_{\nu} = H(a,u) \frac{\sqrt{\pi} e^2}{m_e c} \frac{f \lambda N_{\rm l}}{b}
  \left( 1 - \frac{N_{\rm u} g_{\rm l}}{N_{\rm l} g_{\rm u}} \right)
\end{equation}

where $H(a,u)$ is the Voigt function, $a$ and $u$ being functions of
the frequency $\nu$ and the Doppler width $b$. $e$ and $m_e$ are the
charge and mass of the electron, respectively, $f$ is the oscillator
strength, $\lambda$ the central rest-frame wavelength of the
transition, $N$ the column density, $g$ the statistical weight of the
level, and subscripts l and u indicate the lower and upper level,
respectively, of the transition in question. The flux arriving at a
particular layer is determined at the line centre of the transition.
As long as the absorption line is still in the optically thin regime,
this is equal to integrating the flux over the line profile.  However,
the line-centre flux starts to diverge from the profile-averaged flux
at larger optical depths. The ratio of these fluxes depends on the
optical depth only (and not, as one might expect, on the $b$-parameter
of the absorbing medium), and therefore we can easily convert the flux
at line centre to a profile-averaged flux at any optical depth through
tabulation of this ratio at a range of optical depths.

When considering a cloud with a non-zero size, the layers will be
separated by a distance equal to the cloud size divided by the number
of layers minus one, and the decrease of the flux with distance is
taken into account from layer to layer. We also calculate the flux
decrease due to \hi{} absorption at \lya, \lyb, \lyg, \lyd\ and \lye,
for (excitation) transitions that are near these wavelengths. The
resulting population of the ion levels due to photo-excitation is
calculated as described in \citet{2007A&A...468...83V} and
\citet{2011A&A...532C...3V} (see also
Sect.~\ref{sec:photo-excitation}) for each relevant sublayer.

For ionization, the optical depth is: $\tau = N \sigma$, and we
calculate the sub-layer ionization rate as follows:

\begin{equation}
  R_{\rm sublayer} = \sum\limits_{i=1}^{\rm nshells} \int_{\rm
    E_{th}}^{E_{max}} \frac{\Delta
    t~F_{\nu,i}~\sigma_{\nu,i}~d\nu}{h~\nu}
  \label{eq:ionrate}
\end{equation}

where $\sigma_{\nu, \rm subshell}$ is the cross section (in cm$^2$) of
a particular subshell of an ion. The cross sections of the inner
shells are taken from \citet{Verner93} and \citet{1995A&AS..109..125V}
\footnote{included in the VizieR data base, at
  http://vizier.u-strasbg.fr, with catalog ID: J/A+AS/109/125}, while
the cross sections for the outer shells are taken from
\citet{Verner96}. The integration interval, set by E$_{\rm th}$ and
E$_{\rm max}$, is also defined in these references. The total
ionization rate of each sublayer is obtained by summing the rates of
the different ion subshells. The decrease in the sublayer column
density due to ionization is determined using $dN_{\rm sublayer} =
R_{\rm sublayer} N_{\rm sublayer}$. For ions with excited-level
populations (e.g. \feii{} and \feiii{}), we assume that the ionization
cross section is independent of the excitation level of the ion,
i.e. the fraction of ionizations from each level is considered to be
the same.

When \feii{} is ionized by removal of an inner-shell electron,
i.e. one that is not in the outer 3$d$ or 4$s$ shells, it is generally
not converted to \feiii{}. The vacancy created in the inner shell is
filled by a cascade of radiative (fluorescent) and non-radiative
(Auger) transitions, which can cause additional electrons to be
expelled. The total number of electrons that are removed following
ionization of an electron in a particular shell has been calculated
for various ions and ionization states by \citet{Kaastra93}, whose
results we apply in our program. In practice, practically all \feii{}
ions that are converted to \feiii{} are due to outer shell (3$d$ or
4$s$) ionizations. We also note that although higher-energy photons
(up to several tens of keV) can ionize \feii{}, the majority of
ionizations are caused by lower energy photons (from the ionization
threshold of 16.2~eV up to a few hundred electron-volts). However, the
higher-energy photons become more important with increasing \hi{}
column density, as the hydrogen atoms become more and more effective
at shielding \feii{} and other ions from the lower-energy ionizing
photons. We stress that although we are mentioning \feii{} and
\feiii{} here, the same applies to any other ion that is included,
such as \siii{}, \cii{} and \crii{}.

Since our calculations involve ionization of \feii{} as well as the
population of excited levels of \feiii{}, we need to take into account
that after ionization of \feii{} to \feiii{}, the latter will not
necessarily be in its ground state immediately. Since these are
non-standard calculations and rather important for our modelling of
the \grb\ column densities, they are described in
Sect.~\ref{sec:sigmafe23}.

After a packet of photons ($\Delta t F_{\nu}$) has travelled through
all the layers in the cloud along the sightline, and ionized and
excited ions along its path, the remaining column densities (ground
state and -- if relevant -- excited levels) in each layer are
updated to be used for the next packet coming through. At each time
step, the total ground-state and excited-level column densities are
determined by simply adding up all layer column densities, which can
be compared to measurements.

\end{document}